\documentclass[preprint,amsmath,amssymb,superscriptaddress,showpacs]{revtex4-1}

\usepackage{amsmath,amssymb,graphicx}
\usepackage{enumitem}
\usepackage{hyperref}
\usepackage{amsbsy}
\usepackage{latexsym}
\usepackage{color}
\usepackage{graphicx}
\usepackage{psfrag}
\usepackage[normalem]{ulem}
\usepackage{bm}

\newcommand{\be}{\begin{equation}}
\newcommand{\ee}{\end{equation}}
\newcommand{\bea}{\begin{eqnarray}}
\newcommand{\eea}{\end{eqnarray}}

\newcommand{\comment}[1]{}

\begin{document} 

\title{Equilibrium phases and domain growth kinetics of calamitic liquid crystals}

\author{Nishant Birdi}
\email{srz188382@sire.iitd.ac.in}
\affiliation{School of Interdisciplinary Research, Indian Institute of Technology, Hauz Khas, New Delhi 110016, India.}

\author{Tom L. Underwood}
\email{tlu20@bath.ac.uk}
\affiliation{Department of Chemistry, University of Bath, Bath BA2 7AY, United Kingdom.}

\author{Nigel B. Wilding}
\email{nigel.wilding@bristol.ac.uk}
\affiliation{H.H. Wills Physics Laboratory, University of Bristol, Royal Fort, Bristol BS8 1TL, United Kingdom.}

\author{Sanjay Puri}
\email{purijnu@gmail.com}
\affiliation{School of Interdisciplinary Research, Indian Institute of Technology, Hauz Khas, New Delhi 110016, India.}
\affiliation{School of Physical Sciences, Jawaharlal Nehru University, New Delhi 110067, India.}

\author{Varsha Banerjee}
\email{varsha@physics.iitd.ac.in}
\affiliation{School of Interdisciplinary Research, Indian Institute of Technology, Hauz Khas, New Delhi 110016, India.}
\affiliation{Department of Physics, Indian Institute of Technology, Hauz Khas, New Delhi 110016, India.}

\begin{abstract}
The anisotropic shape of calamitic liquid crystal (LC) particles results in distinct values of energy when the nematogens are placed side-by-side or end-to-end. This anisotropy in energy which is governed by a parameter $\kappa^\prime$ has deep consequences on equilibrium and non-equilibrium properties.  Using the Gay-Berne (GB) model, which exhibits the nematic (Nm) as well as the low temperature \textcolor{black}{smectic (Sm)} order, we undertake large-scale Monte Carlo and molecular dynamics simulations to probe the effect of $\kappa^\prime$ on the equilibrium phase diagram and the non-equilibrium domain growth following a quench in the temperature $T$ or {\it coarsening}. There are two transitions in the GB model: (i) isotropic (I) to Nm at $T_c^1$ and (ii) Nm to \textcolor{black}{Sm} at $T_c^2<T_c^1$. $\kappa^\prime$ decreases $T_c^1$ significantly, but has relatively little effect on $T_c^2$. Domain growth in the Nm phase exhibits the well-known Lifshitz-Allen-Cahn (LAC) law, $L(t)\sim t^{1/2}$ and the evolution is via annihilation of string defects. The system exhibits dynamical scaling that is also robust with respect to $\kappa^\prime$. \textcolor{black}{We find that the Sm phase at the quench temperatures $T$ ($T>T_c^1\rightarrow T<T_c^2$) that we consider has SmB order with a hexatic arrangement of the LC molecules in the layers (SmB-H phase).} Coarsening in this phase exhibits a striking {\it two-time-scale scenario}: first the LC molecules align and develop orientational order (or nematicity), followed by the emergence of the characteristic layering (or smecticity) \textcolor{black}{along with the hexatic bond-orientational-order (BOO) within the layers.} Consequently, the growth follows the LAC law $L(t)\sim t^{1/2}$ at early times and then shows a sharp crossover to a slower growth regime at later times. Our observations strongly suggest that $L(t)\sim t^{1/4}$ in this regime. Interestingly, the correlation function shows dynamical scaling in {\it both} the regimes and the scaling function is {\it universal}. The dynamics is also robust with respect to changes in $\kappa^\prime$, but the smecticity is more pronounced at larger values. Further, the early-time dynamics is governed by string defects, while the late-time evolution is dictated by interfacial defects. \textcolor{black}{We believe this scenario is generic to the Sm phase even with other kinds of local order within the Sm layers.}

\end{abstract}

\maketitle

\section{Introduction}
\label{introduction}

Liquid crystals (LCs) are a state of matter that is intermediate between liquids and crystals as they manifest partial orientational and/or translational order \cite{Stephen_1974,deGennes_1995,Singh_2000,Priestly_2012,Andrienko_2018}. 
The LC {\it mesophases} emerge in response to changes in temperature or concentration. {\it Thermotropic} LCs are usually pure compounds of anisotropic organic molecules which exhibit phase changes by variation of temperature. Depending on their structure, the molecular shape can be rod like (calamitic), disc like (discotic) or banana shaped (bent-core). {\it Lyotropic} LCs are often mixtures of mesogens in a solvent and exhibit phase changes as the concentration of one of the components is varied. Amongst these different types, calamitic LCs are the most well studied due to their simplicity and wide applicability. At high temperatures, they exhibit an {\it isotropic} (I) phase where the rod-shaped molecules are randomly oriented. At low temperatures, the molecules align statistically parallel along a locally preferred axis to yield the {\it nematic} (Nm) phase with purely orientational order. This I-Nm phase transition is first order. If the Nm phase is {\it uniaxial}, it is described by a sign-invariant unit vector known as the {\it director} $\boldsymbol{n}$.

As the temperature is further reduced, some LCs exhibit another transition to the smectic (Sm) phase characterized by partial translational order due to emergence of stacks of layers in addition to the lamellar order along $\boldsymbol{n}$. \textcolor{black}{The Sm nomenclature depends on the ordering in the layers. In the smectic A (SmA) phase, the layers are fluid-like. SmB has local order in the layers: e.g., SmB-H has a six-fold or {\it hexatic} bond-orientational-order (BOO) within the layers and SmB-C has long-ranged translational or {\it crystalline} order in the layers. These smectic phases have been observed experimentally in various LC compounds, either singly or together \cite{Pindak_1979,Goodby_1981,Pindak_1981,Davey_1984,Voronov_2020}. Depending on the coupling between the Nm and the Sm order parameters, the Nm-Sm transitions could be either first order or second order \cite{McMillan_1971,Kralj_2007}.} Several commercially used LCs such as the n-alkyl cyanobiphenyl (nCB) compounds indeed exhibit these twin transitions for $\text{n} \geq 8$, as revealed by light scattering and differential scanning calorimetry experiments \cite{Coles_1979,Chaban_2020}. 

LCs are an important topic of research not only because of their enormous variety of applications, but also because they provide a platform for addressing a variety of fundamental  problems in physics. The Nm phase is extensively used in liquid crystal displays, and the search for newer LCs with improved sensitivity and stability remains an ongoing activity \cite{Chen_2018}. The SmA mesophase provides a general template for {\it striped} systems such as biological membranes and flexible polymer crystallization \cite{Li_2003}. It also shares symmetries with certain types of self-assembled block copolymer films which have applications in photolithography \cite{Harrison_2000,Harrison_2002,Ruiz_2007,Stoykovich_2007}. \textcolor{black}{In recent years, there has been considerable interest in the study of the {\it hexatic} phase which was first predicted as an intermediate state between a crystal and a liquid in the theory of two-dimensional ($2d$) melting \cite{Halperin_1978}. The SmB-H phase provides the 3$d$ analog for the 2$d$ hexatic phase \cite{Davey_1984,Voronov_2020}} Furthermore, LCs are experimentally accessible continuous symmetry systems. They have provided the framework for development of the theory of topological defects \cite{Kleman_2008}. The latter are relevant for a wide range of fields encompassing condensed matter physics, cosmology and biology \cite{Chuang_1991,Pargellis_1991,Lavrentovich_2001,Wang_2016,Kim_2018,Sandford_2020}. 

Experimental measurements to probe various equilibrium and non-equilibrium responses, especially in LC mesophases with lower symmetry, remain a challenge because length scales ($\sim$ nm) of morphologies and timescales ($\sim$ ns) of evolution are often too small to be accessible. Consequently, computer simulations have emerged as a powerful tool for these investigations. In this context, the most important ingredient is the {\it inter-particle potential} which takes into account the anisotropy in the shape of the LC molecules as well as the attractive forces between them. The form proposed by Gay and Berne in 1981, based on the Gaussian overlap model of Berne and Pechukas \cite{Pechukas_1972}, is one of the most popular pair potentials for anisotropic entities \cite{Gay_1981}. The Gay-Berne (GB) model takes into account the aspect ratio $\kappa$ of the mesogens and their energy anisotropy $\kappa^{\prime}$. The latter is defined as the ratio of the potential energies when a pair of mesogens are placed side-by-side (ss) and end-to-end (ee), see Fig.~\ref{figure_1}. Further, the model exhibits I, Nm and Sm phases, and the computed quantities agree well with the corresponding experimental measurements \cite{Luckhurst_1993,Berardi_1993,Bates_1999,Emerson_1994,Cacerez_2014}. The GB model has therefore become a prototype for investigations of LC systems \cite{Adams_1987,Luckhurst_1990,Miguel_1991,Hashim_1995,Satoh_1996,Penna_1996,Cleaver_1996,Neal_1997,Miguel_2002,Zannoni_2001,Haya_2007}.

Laboratory experiments generally require application of external fields that drive the system out-of-equilibrium. The system re-equilibrates, and the approach to equilibrium critically depends on the complexity of the free energy landscape. An important non-equilibrium study in this context is the kinetics of domain growth or {\it coarsening}, initiated by a sudden quench of the system from the disordered phase to the ordered phase \cite{Bray_2002,Puri_2004,Puri_2009}. The domains grow in size via annihilation of defects. The subsequent domain growth, characterized by a growing length scale $L(t)$, is monitored with time. The growth law depends upon several factors such as symmetry of the order parameter, conservation laws, hydrodynamics, etc. It also provides important insights on the barriers to coarsening and relaxation time-scales. Phase ordering in the $d=3$ Nm LCs is well studied using coarse-grained free energy models \cite{Bray_1993,Wickam_1997} and lattice models \cite{Blundell_1992,Zhang_1993_PRE,Bradac_2011,Birdi_2020}. The domain growth obeys the Lifshitz-Allen-Cahn (LAC) law, $L(t) \sim t^{1/2}$ \cite{Allen_1979}, with strings as the dominant defects. However, work in the context of this important non-equilibrium phenomenon for the smectic mesophases remains limited. There have been few experimental \cite{Harrison_2002} and computational \cite{Nasser_2008} studies on coarsening in the $d=2$ SmA phase. They indicated that the orientational correlation length obeys an unusual $L(t)\sim t^{1/4}$ law. Similar growth law, with speculations about logarithmic corrections, has been predicted for the $d=3$ SmA phase using coarse-grained free energy models \cite{Liarte_2015}. Surprisingly, none of these studies address the significant role of the energy anisotropy that is a key feature of calamitic LCs.  

Motivated to augment the above studies, we undertake large-scale simulations of the $d=3$ GB model to understand the consequences of the energy anisotropy on equilibrium and non-equilibrium properties. There are two significant aspects of our study. Firstly, using Monte Carlo (MC) simulations, we identify the phase transition temperatures $T_c^1$ (I$\rightarrow$Nm) and $T_c^2$ (Nm$\rightarrow$Sm) for a range of $\kappa^\prime$ values. These estimates equip us to perform temperature quenches in the Nm and Sm mesophases for the second part of our study. Subsequent to the quench, we study phase ordering kinetics via molecular dynamics (MD) simulations, which are better suited to monitor the systemic evolution as compared to MC simulations. The main results of our paper are as follows:\\
a) The Nm and Sm phases are observed for all values of $\kappa^\prime$. The Nm phase shrinks with increasing values of $\kappa^\prime$ due to a (substantial) decrease in $T_c^1$ and a (marginal) increase in $T_c^2$. \\
(b) When quenched from the I$\rightarrow$Nm phase, domains with orientational order or {\it nematicity} emerge and grow with time. The correlation function $C(r,t)$ vs. $r$ exhibits dynamical scaling indicating the presence of a unique length scale. The scaling function is {\it universal} for different values of $\kappa^\prime$. \\
(c) The tail of the structure factor obeys the generalized Porod law, $S(k,t)\sim k^{-5}$ indicating scattering off {\it string defects.} The growth law in the Nm phase is the usual LAC law, $L(t) \sim t^{1/2}$ characteristic of systems with non-conserved dynamics.\\
(d) {\color{black} For the quenches $T>T_c^1\rightarrow T<T_c^2$, we access the SmB-H phase. The coarsening in this phase is a two stage process: first there is emergence of nematicity, followed by the layering of mesogens or {\it smecticity} along with the development of hexatic order within the layers.} The latter is enhanced by increasing values of $\kappa^\prime$. \\  
(e) The correlation function $C(r,t)$ exhibits dynamical scaling. However the scaling functions show small variations at short distances in the two regimes. These are reflected in the tails of the structure factor: $S(k,t)\sim k^{-5}$ (early time nematicity) indicating scattering off {\it strings}; $S(k,t)\sim k^{-4}$ (late time smecticity) implying scattering off {\it interfaces}. The mechanism of domain growth is thus distinct in the two regimes. The growth law exhibits a previously unreported {\it cross-over} from $t^{1/2}$ to $t^{1/4}$ as time evolves.\\

Our paper is organized as follows. Sec.~\ref{model} provides a detailed discussion of the Gay-Berne (GB) model. In Sec.~\ref{montecarlo}, we present the numerical details and results from our MC simulations for the equilibrium phase diagram of the system for a range of energy anisotropy values. Sec.~\ref{moleculardynamics} presents numerical details and results from our MD simulations for the phase ordering kinetics of the system. The various tools for analyzing the coarsening morphologies are also discussed. In Sec.~\ref{summary}, we conclude with a summary and discussion of our results.

\section{Gay-Berne Model}
\label{model}
The Gay-Berne (GB) model is specially developed to mimic the interactions between ellipsoidal LC molecules, which are of equal size \cite{Gay_1981}. Besides Lennard-Jones (LJ) potential with attractive and repulsive parts which decrease with the intermolecular separation $r$ as $r^{-6}$ and ${r^{-12}}$ respectively, the GB potential includes terms with additional dependence on the orientations of the LC molecules. The potential is anisotropic, and can model the orientational order observed in systems with anisotropic constituents. 

Let us consider two uniaxial LC molecules $i$ and $j$, with orientations defined by the unit vectors ${\boldsymbol{u}_i}$ and ${\boldsymbol{u}_j}$, and centers separated by ${\boldsymbol{r}}$. The GB potential for this prototypical pair is defined by \cite{Gay_1981}:
\begin{equation}
\label{equation_1}
E_{ij}({\boldsymbol{u}_i,\boldsymbol{u}_j,\boldsymbol{r}}) = 4\epsilon({\boldsymbol{u}_i,\boldsymbol{u}_j,\boldsymbol{\hat{r}}}) \left\{\left[\frac{\sigma_0}{r-\sigma({\boldsymbol{u}_i,\boldsymbol{u}_j,\boldsymbol{\hat{r}}})+\sigma_0}\right]^{12} - \left[\frac{\sigma_0}{r-\sigma({\boldsymbol{u}_i,\boldsymbol{u}_j,\boldsymbol{\hat{r}}})+\sigma_0}\right]^6\right\},
\end{equation} 
where $\sigma_0$ scales the distance and ${\boldsymbol{\hat{r}}}$ is the unit vector along ${\boldsymbol{r}}$. The other terms in Eq.~(\ref{equation_1}) are as follows:\\
(a) The orientation-dependent range parameter $\sigma({\boldsymbol{u}_i,\boldsymbol{u}_j,\boldsymbol{\hat{r}}})$ contains information about the shape of the LC molecules and is given by:
\begin{equation}
\label{equation_2}
\sigma({\boldsymbol{u}_i,\boldsymbol{u}_j,\boldsymbol{\hat{r}}}) = \sigma_0\left\{ 1 - \frac{\chi}{2} \left[ \frac{({\boldsymbol{u}_i\cdot\boldsymbol{\hat{r}}+\boldsymbol{u}_j\cdot\boldsymbol{\hat{r}}})^2}{1+\chi({\boldsymbol{u}_i\cdot\boldsymbol{u}_j})}+\frac{({\boldsymbol {u}_i\cdot\boldsymbol{\hat{r}}-\boldsymbol{u}_j\cdot\boldsymbol{\hat{r}}})^2}{1-\chi({\boldsymbol{u}_i\cdot\boldsymbol{u}_j})}\right]\right\}^{-1/2} .
\end{equation}
The shape anisotropy parameter $\chi$ determines the system's capability to form an orientationally ordered phase and is given by: 
\begin{equation}
\label{equation_3}
\chi = \frac{\kappa^2-1}{\kappa^2+1},
\end{equation}
where $\kappa$ is the aspect ratio of the LC molecules. If $\sigma_e$ and $\sigma_s$ are the length and breadth of the molecule, then $\kappa\equiv\sigma_e/\sigma_s$. More precisely, $\sigma_e$ and $\sigma_s$ are the contact distances or intermolecular separations at which the attractive and the repulsive terms in the potential cancel out each other when the LC molecules are in the end-to-end (ee) and the side-by-side (ss) configurations.\\ 
(b) The energy term $\epsilon({\boldsymbol{u}_i,\boldsymbol{u}_j,\boldsymbol{\hat{r}}})$ is defined as: 
\begin{equation}
\label{equation_4}
\epsilon({\boldsymbol{u}_i,\boldsymbol{u}_j,\boldsymbol{\hat{r}}}) = \epsilon_0\epsilon_1^{\mu}\epsilon_2^{\nu},
\end{equation}
where $\epsilon_0$ scales the energy. The parameters $\epsilon_1$ and $\epsilon_2$ are defined follows:
\begin{eqnarray}
\label{equation_5}
\epsilon_1 &=& 1 - \frac{\chi^\prime}{2} \left\{ \frac{({\boldsymbol{\hat{r}}\cdot\boldsymbol{u}_i+\boldsymbol{\hat{r}}\cdot\boldsymbol{u}_j})^2}{1+\chi^\prime({\boldsymbol{u}_i\cdot\boldsymbol{u}_j})}+\frac{({\boldsymbol{\hat{r}}.\boldsymbol{u}_i-\boldsymbol{\hat{r}}\cdot\boldsymbol{u}_j})^2}{1-\chi^\prime({\boldsymbol{u}_i\cdot\boldsymbol{u}_j})}\right\},\\
\label{equation_6}
\epsilon_2 &=& \left[1-\chi^2\left({\boldsymbol{u}_i\cdot\boldsymbol{u}_j}\right)^2\right]^{-1/2}. 
\end{eqnarray}
The energy anisotropy parameter $\chi^\prime$, analogous to Eq.~(\ref{equation_3}), is defined as:
\begin{equation}
\label{equation_7}
\chi^\prime = \frac{\kappa^{\prime1/\mu}-1}{\kappa^{\prime1/\mu}+1},
\end{equation}
where $\kappa^{\prime}$ is the energy anisotropy which if greater than 1 promotes orientational order characteristic of LCs. If $\epsilon_e$ and $\epsilon_s$ are the well depths for the ee and the ss configurations, $\kappa^{\prime}\equiv\epsilon_s/\epsilon_e$. The parameters $\mu$ and $\nu$ modify the well depths of the potential, and hence their impact on the nematicity and smecticity is very subtle. For instance in the ee configuration, $\kappa=3.0$ and $\kappa^\prime=5.0$ yields $\epsilon/\epsilon_0=1/3$ for $\mu=2$, $\nu=1$, and 25/27 for $\mu=1$, $\nu=3$. Similarly in the ss configuration, $\epsilon/\epsilon_0=5/3$ for $\mu=2$, $\nu=1$, and 125/27 for $\mu=1$, $\nu=3$. Fig.~\ref{figure_1} shows variation of the GB potential $E_{ij}/\epsilon_0$ vs. $r_{ij}/\sigma_0$  in the ee configuration for different values of (a) the energy anisotropy $\kappa^\prime$ and (b) the exponents $\mu$, $\nu$. The corresponding variation in the ss configuration is shown in the insets. From the inset of Fig.~\ref{figure_1}(a), it is clear that the ss configuration is energetically favorable for all values of $\kappa^\prime$ {\color{black}(the curves for different values of $\kappa^\prime$ are coincident as the energy in the ss configuration depends only on $\kappa$ and $\nu$).}

Summing up, the GB model contains four essential parameters: $\kappa$, $\kappa^\prime$, $\mu$ and $\nu$. Clearly, there is a large variety of GB homologues which differ from each other in terms of the values chosen for the four parameters \textcolor{black}{\cite{Gay_1981,Berardi_1993,Bates_1999,Zannoni_2001}}. A frequently employed choice of GB parameters is $\kappa=3$, $\kappa^\prime=5$, $\mu=1$ and $\nu=3$ due to Berardi {\it et al.} \cite{Berardi_1993}. The choice of $\kappa=3.0$ is elementary as for real LC systems, the length-to-breadth ratio of the constituent molecules must be equal to or greater than about 3:1. The parameters used by Berardi {\it et al.} have two important features. Firstly, they provides diverse phases, {\it viz.} isotropic, nematic and smectic. \textcolor{black}{The nematic phase is observed over a wide range of temperatures unlike the narrow region observed with the original GB parameterization: $\kappa=3$, $\kappa^\prime=5$, $\mu=2$ and $\nu=1$ \cite{Gay_1981}.} Secondly, the simulation results exhibit convergence with experiments. For instance, the computed temperature variation of the orientational order parameter is in agreement with the experimental data in many real systems \cite{Berardi_1993}. Consequently, the parameters $\kappa=3$, $\kappa^\prime=5$, $\mu=1$ and $\nu=3$  are frequently chosen for simulations of the GB model. \textcolor{black}{We use this set of values in our present work and plan to do a comparative study with other choices of parameters at a later stage.}

\section{Equilibrium studies using Monte Carlo simulations}
\label{montecarlo}
\subsection{Simulation Details}
Prior to investigating the phase ordering kinetics, we first visit the problem of equilibrium phase transitions in the GB model to precisely identify the quench temperatures for the Nm and the Sm regimes. To sample the available phase space, MC simulations are performed in the canonical ($NVT$) ensemble. We use the simulation program DL\_MONTE \cite{Purton_2013,Brukhno_2019} for this purpose. DL\_MONTE is a general-purpose MC program which supports a wide range of MC simulation techniques and interatomic potentials. However, prior to this work it was not applicable to particles with implicit orientations and anisotropic interaction potentials, for instance the GB particles. Our work is a step in that direction since it involved extending DL\_MONTE to make it suitable for such systems. The latest version of DL\_MONTE which includes our improvements is available at \cite{DL_Monte}. This code is a beneficial resource for the research community to study uniaxial GB systems with a number of MC simulation techniques, including grand-canonical and constant-$NPT$ ensembles \cite{Frenkel_2002}. In this work, we only present results using MC in the $NVT$ ensemble. Our DL\_MONTE input files with all the necessary commands, parameters and comments are provided in the Supplementary Material for interested readers together with further details regarding the improvements we have made to DL\_MONTE to treat
the GB and similar models.   

We consider a system of $N$ ellipsoidal particles interacting via the GB potential specified in Eq.~(\ref{equation_1}) and use $\kappa=3.0$, $\kappa^{\prime}=5.0$, $\mu=1.0$ and $\nu=3.0$ unless specified. It is convenient to define scaled variables $E^\ast=E/\epsilon_0= \sum_{i,j;i<j}^N (E_{ij}/\epsilon_0$), $E_1^\ast = E^\ast/N$, $T^\ast=k_BT/\epsilon_0$, $r^\ast=r/\sigma_0$ and $\rho^\ast=N\sigma_0^3/V$. Simulations have been performed on cubic lattices with $N=512$ and $1000$ particles using periodic boundary conditions and $\rho^\ast$ is set to 0.30. The initial configuration is a perfectly aligned state. The MC moves are performed according to the Metropolis algorithm \cite{Frenkel_2002}. In these moves, a randomly selected LC molecule is either translated or rotated with translations and rotations attempted in each move with equal probability. The maximum angle to rotate an LC molecule is kept as $15^\circ$. In order to compute the energy efficiently, a spherical cutoff of $r_c=4.0\sigma_0$ is employed in conjunction with the Verlet neighbor list scheme \cite{Frenkel_2002}. The simulations are performed for $10^6$ MC cycles (one MC cycle corresponds to $N$ MC moves). The initial $5\times10^5$ cycles are necessary for equilibration. The remaining $5\times10^5$ cycles are used for thermal (block) averaging of various thermodynamic quantities of interest and the ensemble averaging is performed over 500 configurations. 

\subsection{Phase Diagram}
The GB phase diagram is obtained by studying the average energy per particle $\langle E_1^\ast \rangle$ vs. $T^\ast$ for values of the energy anisotropy $\kappa^\prime$ = 1.25, 2.5, 5.0, 10.0 and 20.0. For each value, the transition temperatures $T_c^1$ (I$\rightarrow$Nm transition) and $T_c^2$ (Nm$\rightarrow$Sm transition) are identified from the discontinuity in $\langle E_1^\ast\rangle$. It is pertinent to point out here that for our coarsening experiments, we only need approximate boundaries as we consider the quenches far away from these. The I and Nm phases are confirmed by evaluation of the orientational order parameter, $\mathcal{S}$ given by:
\begin{equation}
\label{equation_8}
\mathcal{S} = \langle P_2\left(\cos\theta_{i}\right)\rangle = \left\langle \frac{3\cos^2\theta_{i}-1}{2}\right\rangle,
\end{equation}
where $\cos\theta_i = \boldsymbol{u}_i\cdot \boldsymbol{n}$ and the angular brackets $\langle\cdot\cdot\cdot\rangle$ indicate an ensemble average. $\mathcal{S}=0$ in the I phase, while $\mathcal{S}=1$ in the perfectly aligned Nm phase. \textcolor{black}{The defects correspond to regions with $\mathcal{S} \simeq 0$, even if the defect cores are biaxial \cite{Pelcovits_2006}.} 

\textcolor{black}{The smectic (layered) phases can be distinguished from the I and Nm phases by evaluating the translational order parameter \cite{Bates_1999}:
\begin{equation}
\label{equation_9}
\mathcal{T}=\lvert\langle\tau(\boldsymbol{r_i})\rangle\rvert=\lvert\left\langle\exp(2i\pi\boldsymbol{r_i}\cdot\boldsymbol{n}/d_l)\right\rangle\rvert, 
\end{equation}
where $\boldsymbol{r_i}$ is the  position vector of the LC molecule $i$ and $d_l$  is the layer spacing. In simulations, $\mathcal{T}$ is determined by first separately performing the ensemble averaging of the real and imaginary terms: $\cos(2i\pi\boldsymbol{r_i}\cdot\boldsymbol{n}/d_l)$ and $\sin(2i\pi\boldsymbol{r_i}\cdot\boldsymbol{n}/d_l)$, followed by calculation of the modulus for the ensemble average and further maximizing it with respect to $d_l$ \cite{Bates_1999}. In the I phase, $\mathcal{T}$ $\rightarrow$ 0 as the layers are not well defined. On the other hand, $\mathcal{T}$ $\rightarrow$ 1 in a perfectly layered structure.}
   
\textcolor{black}{A clear distinction between the SmA and SmB phases can be made by evaluating the hexatic bond-orientational order parameter given by \cite{Bates_1999}:
\begin{equation}
\label{equation_10}
\mathcal{C}_6=\lvert\langle\psi_6(\boldsymbol{r_i})\rangle\rvert=\left\lvert\left\langle\left(\frac{\sum_kw(r_{ik}^\ast)\exp(6i\phi_{ik}))}{\sum_kw(r_{ik}^\ast)}\right)\right\rangle\right\rvert,
\end{equation}
where the summation is over the nearest neighbours (nn) $k$ of the LC molecule $i$, $\phi_{ik}$ is the angle between the vector ($\boldsymbol{r_i}-\boldsymbol{r_k}$) projected onto the plane normal to the director and a fixed reference axis ($x$ axis say) and $w(r_{ik}^\ast)$ is a cutoff function to select the nn for evaluation of $\psi_6(\boldsymbol{r_i})$. (Like $\mathcal{T}$, for $\mathcal{C}_6$ also first the ensemble averaging is done separately for the real and imaginary terms followed by calculation of the modulus for the average.) It is important to use the cutoff function as the number of nn might not be 6 and could be 7, 5 or 4 when the local translational order is imperfect. In our work, we used the procedure in reference \cite{Bates_1999} to evaluate this cutoff function: $w(r_{ik}^\ast)$ is unity for $r_{ik}^\ast$ below 1.4, zero for $r_{ik}^\ast$ above 1.8 and with a linear interpolation in between these two extremes. If there is hexatic order, $\mathcal{C}_6$ has an appreciable non-zero value and also the ensemble average for the cutoff function, $\left \langle w(r_{ik}^\ast) \right \rangle \rightarrow 6$. In it's absence, $\mathcal{C}_6$ vanishes and $\left \langle w(r_{ik}^\ast) \right \rangle < 6$.}

The radial distribution function $g(r)$ is routinely obtained in scattering experiments, and measures the probability of finding two molecules separated by distance $r$ relative to that in an ideal gas \cite{Frenkel_2002}. It is a useful tool to distinguish between the local order in the \textcolor{black}{different phases \cite{Bates_1999}, and is given by:}
\begin{equation}
\label{equation_11}
g(r^\ast) = \frac{\langle \overline{\rho(r^\ast)}\rangle}{\rho_0},
\end{equation}
where $\rho_0=N/V$ is the density of the ideal gas and $\overline{\rho(r^\ast)}$ is the average density of the system around $r^\ast$. The numerical evaluation is facilitated by the following formula \cite{Frenkel_2002,Billeter_1998}: 
\begin{equation}
\label{equation_12}
g(r^\ast)=\frac{1}{N\rho_0}\bigg\langle\sum_{\stackrel{i,j}{i\neq j}}^N\frac{\delta(r^\ast-r_{ij}^\ast)}{(4/3)\pi[(r^\ast+\Delta r^\ast)^3-(r^\ast)^3]}\bigg\rangle.
\end{equation}
The $\delta$ function is unity if $r_{ij}^\ast$ falls within the shell centered on $r^\ast$ and is zero otherwise.  The division by $N$ is done to normalize $g(r^\ast)$ to a per-molecule function. By construction, $g(r^\ast)=1$ for an ideal gas and any deviation implies correlations between the particles due to the intermolecular interactions. In the Nm phase, it has a noticeable maximum for small-$r^\ast$ and is 1 for large-$r^\ast$ indicating short-range order. \textcolor{black}{In the SmA phase, $g(r^\ast)$ exhibits a large nn peak and small oscillations thereafter with periodicity $\sim d_l$ due to the layering. In the SmB-H phase on the other hand, $g(r^\ast)$ shows an oscillatory behaviour with many sharp peaks separated by $r^*\simeq 1$ and a split-peak at $r^*\simeq 2$ characteristic of {\it hexatic} order within the layers. In the case of SmB-C order, the peak intensities do not decay till large distances.} 

We now proceed to evaluate the phase diagram in $(T^\ast,\kappa^\prime)$-space for $\kappa^\prime$ = 1.25, 2.5, 5.0, 10.0 and 20.0. In Fig.~\ref{figure_2}(a), we show the variation of the per-particle equilibrium energy $\langle E_1^\ast\rangle$ vs. $T^\ast$ for $N=512$ (open up triangles) and $N=1000$ (open down triangles) at $\kappa^\prime=5.0$. (For brevity, we do not show the data sets for other values of $\kappa^\prime$.) The angular brackets indicate thermal averages. The simulation data coincide, indicating that the average energies are independent of the system size. We also plot the corresponding benchmarking results of Berardi {\it et al.} for $N=512$ (white circles) and $N=1000$ during heating (white up triangles) and cooling (white down triangles) protocols \cite{Berardi_1993}. There is an excellent agreement with our simulation results obtained using DL\_MONTE, even though the initial conditions in the two sets of simulation experiments are quite distinct. Furthermore, the effects of system size on the average energy are negligible (except near the transition in some cases). The left and right edges of the Nm region (in green) provide the scaled transition temperatures $T^{\ast 2}_c$ and $T^{\ast 1}_c$ as indicated. Both the transitions are discontinuous (first-order). 

\textcolor{black}{To confirm the phase corresponding to each data point, we have also evaluated for the $N=1000$ system, the temperature variation of the average order parameters $\langle \mathcal{S}\rangle$, $\langle \mathcal{T}\rangle$ and $\langle \mathcal{C}_6\rangle$, or $g(r^\ast)$ vs. $r^\ast$ as appropriate. Fig.~\ref{figure_2}(b) shows the typical behaviour of the order parameters in various phases. Note the rise in $\langle \mathcal{S}\rangle$ as the temperature crosses $T^{\ast 1}_{c}$ from above, indicating the onset of nematic order. Similarly, note the rise in translational order parameter $\langle \mathcal{T}\rangle$ as the temperature crosses $T^{\ast 2}_{c}$ indicating the development of SmA order (while evaluating $\mathcal{T}$, we have observed that the layer spacing $d_l$ which maximizes $\mathcal{T}$ is less than the aspect-ratio $\kappa=3.0$ due to interdigitation of the layers \cite{Bates_1999}). At slightly lower values of $T$, the hexatic order parameter $\langle \mathcal{C}_6\rangle$ becomes significant suggesting that the phase is SmB-H. 
Accurate determination of the phase boundaries will require careful evaluations. We refrain from going in this direction as the focus of the present study is on coarsening and approximate phase boundaries are sufficient for this purpose. (We wish to mention here that though the equilibrated configurations, their energies in various phases and the order parameter $\langle \mathcal{S}\rangle$ were obtained using DL\_MONTE, the codes for evaluation of the order parameters $\mathcal{T}$ and $\mathcal{C}_6$ were written separately.)} 

\textcolor{black}{To evaluate $g(r^\ast)$, we have used a set of $745$ bins with a separation cut-off of $r^\ast=7.45$} which yields $\Delta r^\ast=0.01$ (see Supplementary Material for details regarding specification of parameters for DL\_MONTE). Fig.~\ref{figure_2}(c) shows the $g(r^\ast)$ vs. $r^\ast$ behaviour for the \textcolor{black}{Nm phase (green line, $T^\ast=2.5$) and the Sm phase (blue line, $T^\ast=1.0$). It is characterized by multiple sharp peaks with a split peak at $r^\ast\simeq2$, characteristic of the SmB-H order. Further, it can be seen that $g(r^\ast)$ decays as $r^\ast$ increases, implying that the translational order is lost at larger distances and hence the SmB phase is not of crystal type.}

Fig.~\ref{figure_2}(d) depicts the variation of $T^{\ast 1}_c$ (white up triangles) and $T^{\ast 2}_c$ (white down triangles) for a range of $\kappa^\prime$ values. To be noted here is that $T^{\ast 1}_c$ decreases considerably with increasing $\kappa^\prime$ thereby shrinking the Nm phase. On the other hand, $T^{\ast 2}_c$ increases only slightly. \textcolor{black}{With the original GB parameterization, de Miguel {\it et al.} also reported similar observations \cite{Miguel_1996}.} 

\section{Kinetic properties from Molecular Dynamics Simulations}
\label{moleculardynamics}
\subsection{Methodology}
\label{methodology}
We now turn to a study of domain growth kinetics in the $d=3$ GB model via MD simulations. To the best of our knowledge, they have not been addressed. There have been some coarsening studies for the Nm phase of the GB model in $d=3$. A relevant contribution in this context is by Billeter {\it et al.} \cite{Billeter_1999}. They observed the usual defect structures (e.g. disclinations and monopoles), but could not extract a reliable growth law due to the small system size considered ($\sim$60000 particles). Using large systems ($\sim$260000 particles), we perform deep quenches from the I phase to the Nm and the \textcolor{black}{Sm} phases [shown in Fig.~\ref{figure_2}(d)] and allow the system to evolve for long times. Our main interest is to determine the novel defects and growth laws in the \textcolor{black}{Sm} phase, and the impact of the energy anisotropy $\kappa^\prime$ on ordering. We also undertake analogous studies for the Nm phase to complement previous works \cite{Billeter_1999}.  

All our MD simulations have been performed in the $NVT$ ensemble using the LAMMPS software package \cite{Plimpton_1995,Lammps}. The details regarding implementation of the model into LAMMPS \footnote{Note that LAMMPS technically implements the generalized Gay-Berne model \cite{Berardi_1995} in which interactions of the ellipsoidal particles are characterized via a (diagonal) shape matrix $S=$ diag($\sigma_a,\sigma_b,\sigma_c$) and a (diagonal) energy matrix $E=$ diag($\epsilon_a,\epsilon_b,\epsilon_c$), where $\sigma_a,\sigma_b,\sigma_c$ are the lengths and $\epsilon_a,\epsilon_b,\epsilon_c$ are the relative well depths of interaction along the three semi-axes of an ellipsoid. The GB model discussed in Sec.~\ref{model} and employed in Sec.~\ref{montecarlo} was the potential originally presented for ellipsoidal particles of equal size. The generalized GB model was developed to represent dissimilar biaxial ellipsoids \cite{Berardi_1995}. It reduces to the GB model described in Sec.~\ref{model} when the molecules become uniaxial. Hence, for our study, we work with the generalized GB model in the uniaxial limit.} and also the analytical expressions for the forces and torques, have been described in \cite{Brown_2009}.
We consider $N=262144$  uniaxial ellipsoidal particles confined in a cubic box of linear size $L_s\sigma_0$ with periodic boundary conditions in all three coordinate directions. Hence the volume $V=$ ($L_s\sigma_0$)$^3$ = ($95.6033\sigma_0$)$^3$ such that $\rho^\ast\simeq 0.3$. The parameters in the GB potential of Eq.~(\ref{equation_1}) are those employed by Berardi {\it et al.} \cite{Berardi_1993}.  We also study the effects of varying the energy anisotropy parameter $\kappa^\prime$. The MD runs are carried out using the standard velocity Verlet algorithm. In LAMMPS, the dimensionless MD time unit $t_0=\sqrt{m_0\sigma_0^2/\epsilon_0}=1.0$. We choose the reduced MD integration time step $\Delta t^\ast=\Delta t/t_0 = 0.001$.  The temperature $T^\ast$ is controlled and maintained constant via the Nos\'e-Hoover thermostat, which is known to preserve hydrodynamics \cite{Frenkel_2002,Nose_1984,Binder_2004}. The homogeneous initial configurations are prepared by equilibrating the system at a high temperature ($T=6.0$) for about $10^5$ MD steps. To initiate the coarsening process ($t=0$), the system is quenched to the indicated temperatures in Fig.~\ref{figure_2}(b). The evolution of the system is then monitored. All statistical quantities of interest are averaged over $19$ independent initial conditions. Our input files and parameters used in LAMMPS are provided in the Supplementary Material.   

\subsection{Characterization Tools}
\label{cztools}
For a translationally invariant system, the usual probe to characterize configurational morphologies is the equal-time correlation function \cite{Puri_2009}: 
\begin{equation}
\label{equation_13} 
C(\vec{r},t)=\langle\psi(\vec{r_1},t)\psi(\vec{r_2},t)\rangle-\langle\psi(\vec{r_1},t)\rangle\langle\psi(\vec{r_2},t)\rangle ,
\end{equation}
where $\psi(\vec{r},t)$ is a suitable order parameter, $\vec{r}=\vec{r_2}-\vec{r_1}$ and $\langle \cdot\cdot\cdot \rangle$ represents the ensemble average. Small-angle scattering experiments yield the structure factor:
\begin{equation}
\label{equation_14}
S(\vec{k},t) = \int d\vec{r}~e^{i \vec{k} \cdot \vec{r}}~C(\vec{r},t), 
\end{equation}
where $\vec{k}$ is the wave-vector of the scattered beam. A characteristic length scale $L(t)$ is usually defined as the distance at which $C(\vec{r},t)$ decays to, say, $0.2$ times its maximum value. If the domain growth is characterized by a unique length scale $L(t)$, then $C(\vec{r},t)$ and $S(\vec{k},t)$ show the dynamical scaling property  \cite{Puri_2004,Puri_2009}: $C(\vec{r},t) = g(r/L)$; $S(\vec{k},t) = L^d f(kL)$. The asymptotic (large-$k$) tail of $S(\vec{k},t)$ contains information about the defects in the system. Continuous $O(n)$ spin models exhibit the {\it generalized Porod law}, with the asymptotic form: $S(k,t) \sim k^{-(d+n)}$ \cite{Porod_1982,Yono_1988,Bray_1991}. For $n=1$, the defects are interfaces, and the corresponding scattering function exhibits the usual {\it Porod law}: $S(k,t) \sim k^{-(d+1)}$. For $n>1$, the different topological defects are vortices ($n=2,\ d=2$), strings ($n=2,\ d=3$), and monopoles or hedgehogs ($n=3,\ d=3$). So in $d=3$, $S(k,t)\sim k^{-5}$ or $\sim k^{-6}$ depending on whether strings or monopoles dominate in the defect dynamics.

In LC mesophases such as the Sm phase, an appropriate measure of the orientational and translational order is provided by the longitudinal pair correlation function $g_{\parallel}(r_{\parallel})$ (parallel to the long axis of the LC molecules) and the transverse pair correlation function $g_{\perp}(r_\perp)$ (perpendicular to the long axis of the LC molecules) \cite{Richter_2006}. Evaluation of  $g_{\parallel}(r_{\parallel})$ employs a cylindrical volume to probe the LC molecules aligned in the ee configuration and is given by \cite{Richter_2006,Guzman_2014}:
\begin{equation}
\label{equation_15}
g_\parallel(r_\parallel)=\bigg\langle\sum_{i\neq j}^N\frac{\delta(r_\parallel-r_{ij,\parallel_i}) \theta(\sigma_0-r_{ij,\perp_i}) \theta(\sigma_0-r_{ij,\perp_j})}{N\rho^\ast\pi(\sigma_0/2)^2h}\bigg\rangle.
\end{equation}
The Heaviside step function $\theta(x)=1$ when $x\geq0$ and $\theta(x)=0$ otherwise, $\langle\cdot\rangle$ indicates an ensemble averaging over different initial (independent) conditions, $h$ is the cylinder height used to discretize the volume, $r_{ij,\parallel_i}=|\boldsymbol{r}_{ij,\parallel_i}|=|\boldsymbol{r}_{ij}\cdot\boldsymbol{u}_i|$ is the center-of-mass separation along the director of molecule $i$ (the director for molecule $j$ of the pair could also be considered in this $\delta$-function evaluation as the value remains almost the same in the ee configuration). 
$r_{ij,\perp_{i}}=|\boldsymbol{r}_{ij,\perp_{i}}|=|\boldsymbol{r}_{ij}-\boldsymbol{r}_{ij,\parallel_{i}}|$ and $r_{ij,\perp_{j}}=|\boldsymbol{r}_{ij,\perp_{j}}|=|\boldsymbol{r}_{ij}-\boldsymbol{r}_{ij,\parallel_{j}}|$ are the corresponding transverse separations from $\boldsymbol{u}_i$ and $\boldsymbol{u}_j$. The quantity $g_\parallel(r_\parallel)$ probes the average orientation of the LC molecules and the layering or smecticity in the system. 

Evaluation of $g_\perp(r_\perp)$ employs hollow, concentric cylinders to probe LC molecules aligned in the ss configuration and is given by \cite{Richter_2006,Guzman_2014}:
\begin{equation}
\label{equation_16}
g_\perp(r_\perp)=\bigg\langle\sum_{i\neq j}^N\frac{\delta(r_\perp-r_{ij,\perp_i}) \theta(\delta L_\perp/2-r_{ij,\parallel_i}) \theta(\delta L_\perp/2-r_{ij,\parallel_j})}{N\rho^\ast\pi((r_\perp+\delta L_\perp)^2-r_\perp^2)h}\bigg\rangle.
\end{equation}
In the above equation, $\delta L_\perp$ represents the thickness of the hollow cylinder and as for the $g_\parallel(r_\parallel)$ case, we can use the director for molecule $j$ of the pair in the $\delta$-function evaluation since the value remains almost the same in the ss configuration. The quantity $g_\perp(r_\perp)$ probes the translational structure about the LC molecules and their arrangement within layers.

\subsection{Morphologies, Textures and Growth Laws}
\label{growthlaws}
\begin{figure}[ht]
\includegraphics[width=0.76\textwidth]{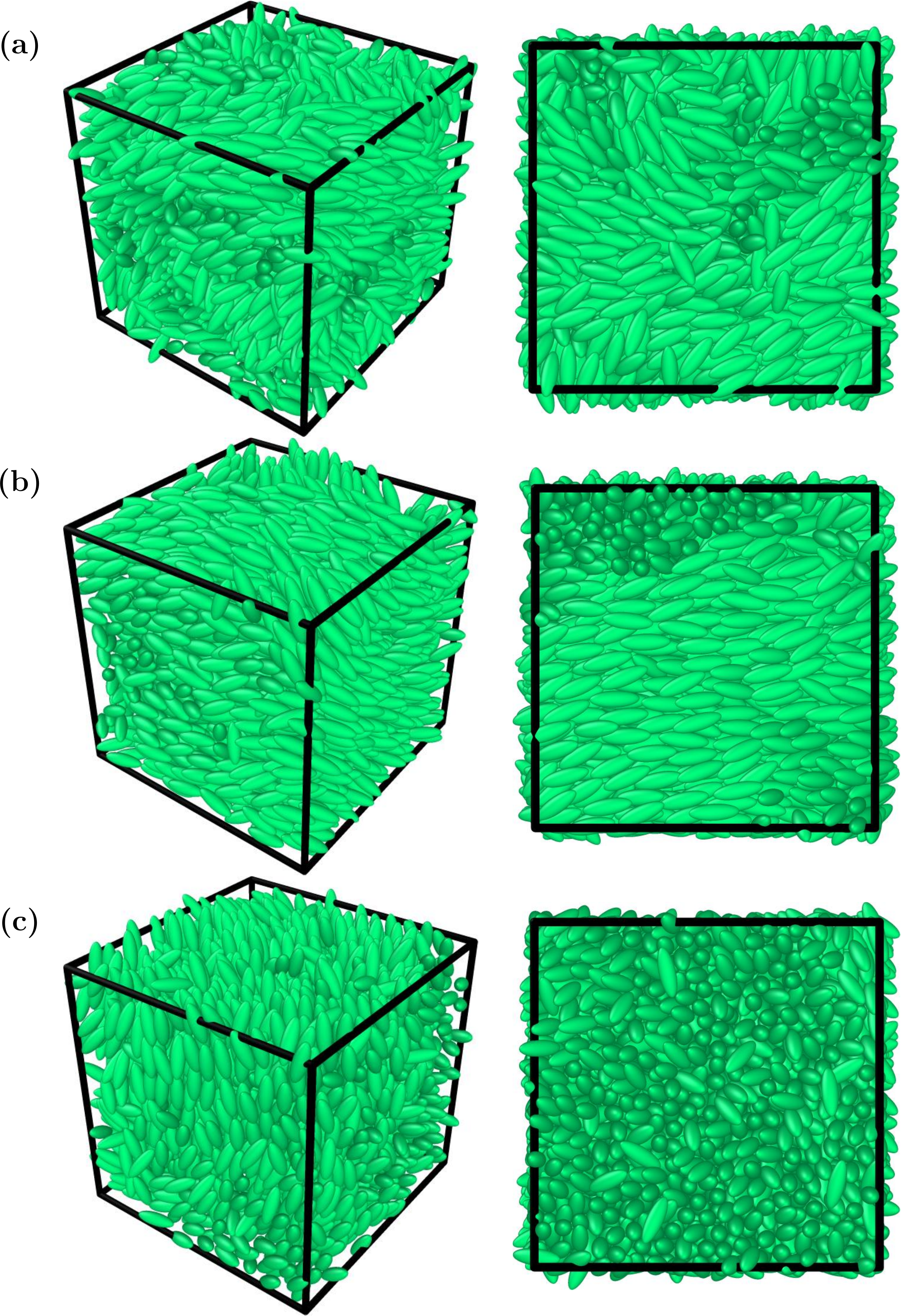}
\caption{Evolution snapshots after after a quench from the I phase to the Nm phase ($T^\ast = 2.5$) for different values of $t$ and $\kappa^\prime$: (a) $\kappa^\prime=5.0$, $t=8192$, (b) $\kappa^\prime=5.0$, $t=53248$ and (c) $\kappa^\prime=10.0$, $t=53248$. 
Left: $20^3$ corner portions of the entire simulation box allowing for visibility of individual particles. Right: the corresponding top $d=2$ cross-sections.}
\label{figure_3}
\end{figure}

Let us first discuss the kinetics of domain growth following  a quench from the I phase to a temperature $T^\ast=2.5$ in the Nm phase. The left panel of Fig.~\ref{figure_3}, shows representative snapshots from the time evolution of the configurational structure for (a) $\kappa^\prime=5.0$, $t = 8192$; (b) $\kappa^\prime=5.0$, $t=53248$; and (c) $\kappa^\prime=10.0$, $t=53248$. For clarity, we have shown only a $20^3$ corner of the entire box. These corners on average consist of about $N=2400$ particles. The right panel shows the corresponding top surface ($d=2$ cross-sections). There is emergence and growth of orientational Nm order, and the energy anisotropy does not affect the coarsening phenomenon. We characterize the morphologies and their texture by evaluating the correlation function and the structure factor. These are obtained by a coarse-graining procedure in which the system is divided into non-overlapping sub-boxes of size ($3.0\sigma_0$)$^3$. The sub-box size is carefully chosen to ensure that each one contains about 8 to 10 particles. The continuum LC configurations are thus mapped onto a simple cubic lattice of size ($32\sigma_0$)$^3$. The relevant order parameter is the orientational order parameter defined in Eq.~(\ref{equation_8}), i.e., $\overline{P_2(\cos\theta_i)}$. Here $\theta_i$ is the angle made by a molecule located in the $i$-th box with the global director $\boldsymbol{n}$ of the system (determined by taking average of orientations for all the $N$ particles present in the system), and the overline implies an average over all the particles in the sub-box. 

\begin{figure}[ht]
\includegraphics[width=1.0\textwidth]{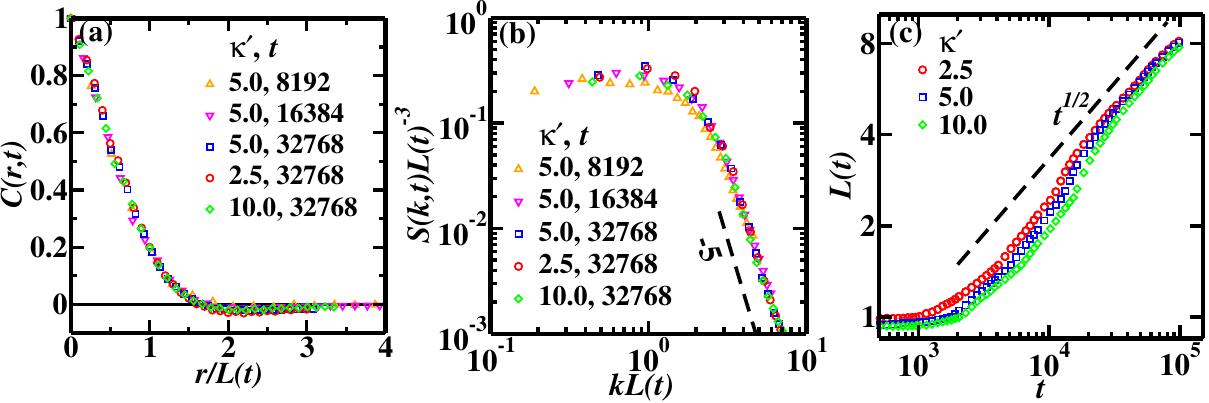}
\caption{(a) Scaled correlation functions $C(r,t)$ vs. $r/L(t)$ for the Nm phase at the specified values of $\kappa^\prime$ and $t$. The data collapse indicates that the system exhibits dynamical scaling, and the morphologies are invariant with respect to $\kappa^\prime$ and $t$ apart from a scale factor. (b) Corresponding scaled structure factors $S(k,t)L(t)^{-3}$ vs. $kL(t)$ on a log-log plot. The dashed line of slope -5 denotes the expected generalized Porod tail. (c) Domain length scale $L(t)$ vs. $t$ for different values of $\kappa^\prime$. The growth obeys the LAC law, $L(t)\sim t^{1/2}$ (dashed line) characteristic of systems with non-conserved order parameter and is slower for larger values of $\kappa^\prime$.}
\label{figure_4}
\end{figure}

Fig.~\ref{figure_4}(a) shows the scaled correlation function, $C(r,t)$ vs. $r/L(t)$ at $t$ = 8192, 16384 and 32768 for $\kappa^\prime$ = 5.0 and at $t=32768$ for $\kappa^\prime$ = 2.5 and 10.0. The data exhibits {\it dynamical scaling} as well as {\it super-universality} with respect to the energy anisotropy $\kappa^\prime$. The dynamical scaling property demonstrates that the coarsening patterns are statistically self-similar in time. The property of super-universality indicates that the morphologies in the Nm phase are independent of the energy anisotropy. \textcolor{black}{This is expected because the GB energy between a pair of ellipsoids in the ss arrangement depends only on $\kappa$ and $\nu$.} A log-log plot of the corresponding scaled structure factors, $S(k,t)L(t)^{-3}$ vs. $kL(t)$ is shown in Fig.~\ref{figure_4}(b). In the asymptotic large-$k$ limit, the structure factor follows the generalized Porod law: $S(k,t)\sim k^{-5}$, indicating scattering off string defects \cite{Bray_1993,Birdi_2020}. Next, we study the time-dependence of the domain size. Fig.~\ref{figure_4}(c) shows the variation of $L(t)$ vs. $t$ on a log-log scale for $\kappa^\prime$ = 2.5, 5.0 and 10.0 respectively. It can be clearly observed that after an initial transient, the evolving systems are consistent with the $t^{1/2}$ growth regime (LAC law) characteristic of systems with non-conserved order parameter. 
The system size and time scales of our simulation are sufficient to establish the LAC domain growth law for the Nm phase of the GB model although there is onset of finite-size effects at late times.  

We now come to the primary focus of our paper: kinetics of domain growth in the Sm mesophase. The coarsening is initiated by a quench from $T^\ast=6.0$ (I phase) to $T^\ast = 1.0$ (Sm phase). Fig.~\ref{figure_5} shows the prototypical evolution morphologies for: (a) $\kappa^\prime=5.0$, $t = 4096$; (b) $\kappa^\prime=5.0$, $t=98304$; and (c) $\kappa^\prime=10.0$, $t=98304$. As in Fig.~\ref{figure_3}, we have shown only a $20^3$ corner of the entire box, consisting of about 2400 particles on average. The frames on the right show the corresponding top surface. It is interesting to note the initial onset of nematicity at the earlier time ($t=4096$) followed by the development of smecticity at later time ($t=98304$). Additionally, as observed in Fig.~\ref{figure_5}(c), the smecticity is significantly enhanced for $\kappa^\prime=10.0$. 

\begin{figure}[ht]
\includegraphics[width=0.45\textwidth]{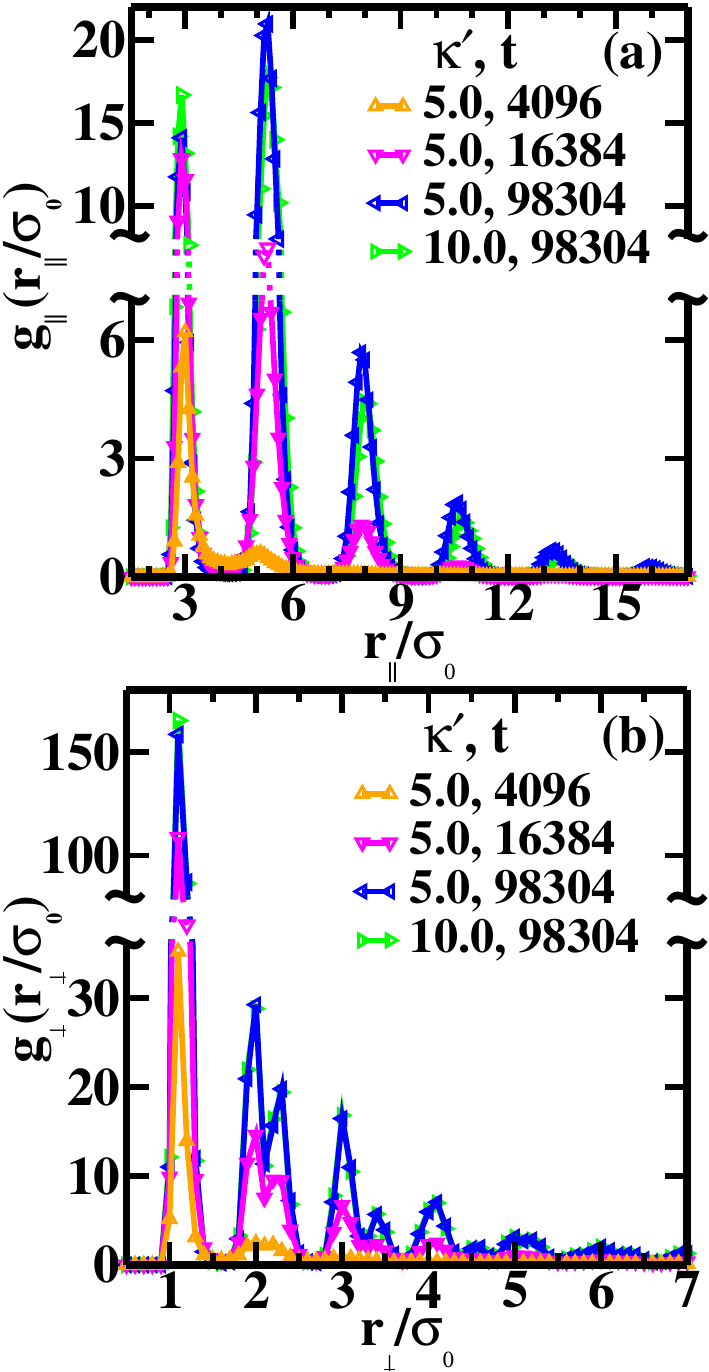}
\caption{(a) Time evolution of the longitudinal radial distribution functions $g_\parallel(r_\parallel/\sigma_0)$ vs. ($r_\parallel/\sigma_0$) for the \textcolor{black}{I$\rightarrow$Sm quench at the} specified values of $\kappa^\prime$ and $t$ signaling layering in the system which occurs only at later times ($t\gtrsim 10^4$). As time evolves, subsequent peaks emerge at separations corresponding to multiples of the ellipsoid length ($3\sigma_0$) indicating ee alignment for the LC molecules \textcolor{black}{(the separations between the peaks is $\lesssim3\sigma_0$ due to interdigitation of the layers).} With increase in $\kappa^\prime$, the intensity of the first peak increases and the variation between the intensities for the first and second peaks decreases. Hence, the degree of smecticity increases as $\kappa^\prime$ increases. (b) Time evolution of the corresponding  transversal radial distribution functions $g_\perp(r_\perp/\sigma_0)$ vs. ($r_\perp/\sigma_0$). These functions indicate alignment within the layers. \textcolor{black}{The peaks in these functions indicate SmB-H ordering, see corresponding text for details.}}
\label{figure_6}
\end{figure}

To characterize the Sm order, Fig.~\ref{figure_6} shows for the specified values of $t$ and $\kappa^\prime$: (a) the longitudinal pair correlation function $g_\parallel(r_\parallel/\sigma_0)$ vs. $r_\parallel/\sigma_0$ evaluated using Eq.~(\ref{equation_11}) and (b) the transversal pair correlation function $g_\perp(r_\perp/\sigma_0)$ vs. $r_\perp/\sigma_0$ evaluated using Eq.~(\ref{equation_12}). Notice that in Fig.~\ref{figure_6}(a), the early-time behaviour at $t=4096$ predominantly exhibits a single peak at $3\sigma_0$ - the length of the LC molecule. It is characteristic of Nm order with molecular alignment along an average direction or the director. As time evolves, there is emergence of newer peaks at approximately $6\sigma_0$, $9\sigma_0$, etc. as the system coarsens. This signifies development of long-range longitudinal order or layers with an inter-layer spacing of $\sim3\sigma_0$. \textcolor{black}{(The average separation between the peaks can be an approximate measure of the inter-layer spacing $d_l$. In  Fig.~\ref{figure_6}(a), $d_l\approx  2.5\sigma_0$ due to interdigitation of the neighbouring layers.)} 
Notice that increase in $\kappa^\prime$ reduces the intensity variation between the first and second peaks implying enhancement of smecticity. (Recall that perfect translational order is characterized by peaks of equal intensity in the pair distribution function). On the other hand, $g_\perp(r_\perp/\sigma_0)$ vs. $r_\perp/\sigma_0$ in Fig.~\ref{figure_6}(b) exhibits peaks around multiples of $\sigma_0$ - the width of the ellipsoidal LC molecule.
\textcolor{black}{The splitting of the peak at $r_\perp\simeq 2.0\sigma_0$, a signature of BOO within the layers, is clearly seen at later times. Further, the translational order within the layers is not long-ranged as consecutive peaks have decreasing intensity. These characteristics, along with the non-zero value of the hexatic order parameter $\langle \mathcal{C}_6\rangle$ in Fig.~\ref{figure_2}(b), confirm the presence of the SmB-H phase. The intra-layer BOO is not affected by $\kappa^\prime$.} 

The observations from Fig.~\ref{figure_6} suggest the following scenario for coarsening of the \textcolor{black}{SmB-H} phase: it is a {\it two-timescale} process, with the onset of Nm order followed by \textcolor{black}{SmB-H} order. \textcolor{black}{To confirm this, we have also performed a similar study for the Nm$\rightarrow$SmB-H quenches and evaluated these distribution functions. Except for the emergence of multiple peaks at earlier time in the longitudinal pair correlation function (it's now a one-timescale process as Nm order is already present and hence with time, the system exhibits only layering with BOO), the behavior of the correlation functions is qualitatively similar to that observed in Fig.~\ref{figure_6} and hence we do not present them separately. Quantitatively, the peaks in both the functions have much higher intensity compared to the corresponding peaks in Fig.~\ref{figure_6}, indicating {\it faster} layering as well as development of BOO. We do not present these data sets to prevent repetition. The {\it two-time-scale} process is the most significant outcome from our study of the GB model. It is further reiterated by the growth laws which will be discussed shortly.}

\begin{figure}[ht]
\includegraphics[width=1.0\textwidth]{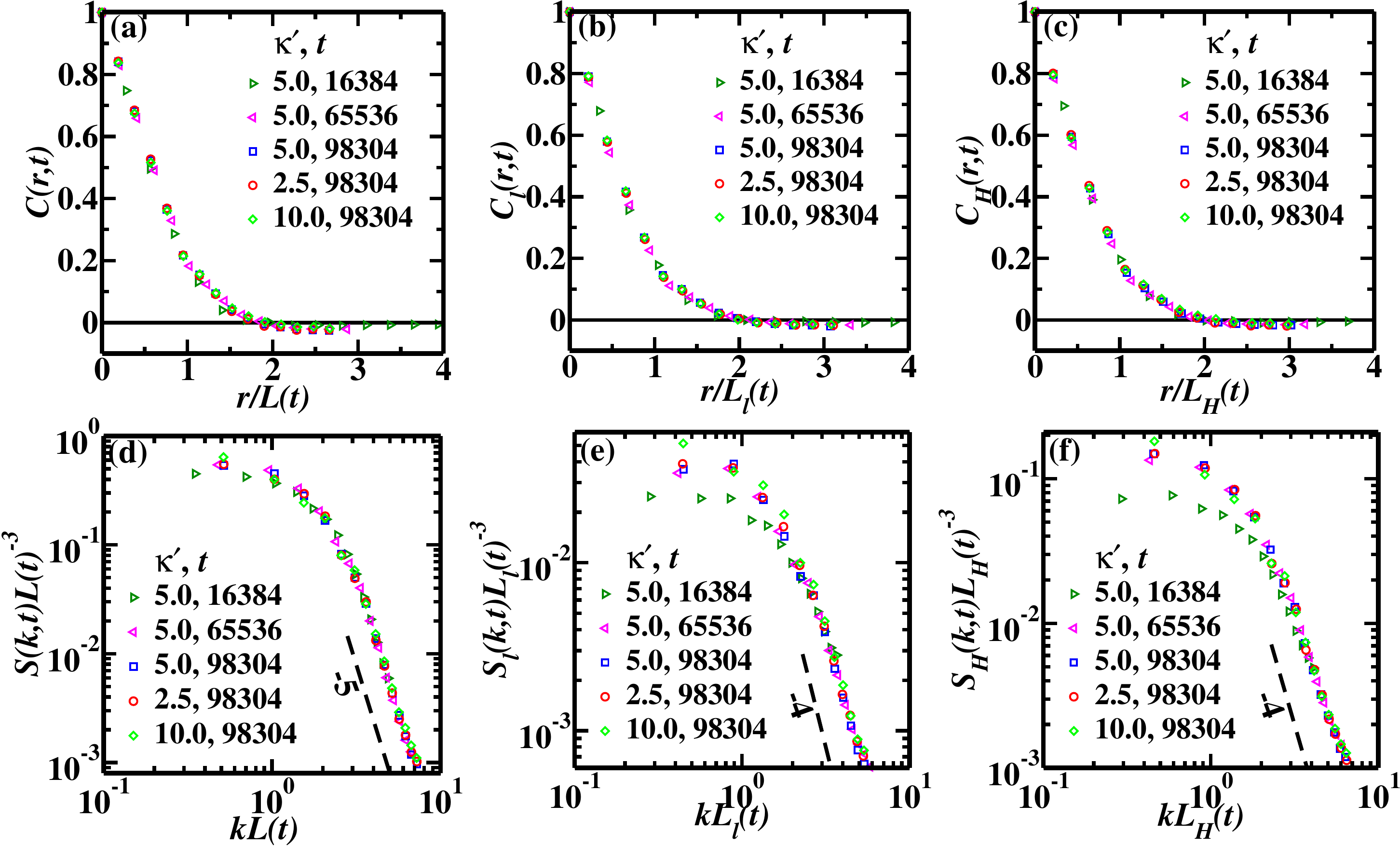}
\caption{\textcolor{black}{Scaled correlation functions for the SmB-H phase at specified values of $\kappa^\prime$ and $t$: (a) $C(r,t)$ vs. $r/L(t)$, (b) $C_l(r,t)$ vs. $r/L_l(t)$ and (c) $C_H(r,t)$ vs. $r/L_H(t)$ (see text for evaluation of these correlation functions). The data collapse in all the three evaluations suggests that the system exhibits dynamical scaling, and the morphologies are invariant with respect to $t$ and $\kappa^\prime$ although the coarsening scenario is distinct at early and late times. The corresponding structure factors are shown on a log-log plot: (d) $S(k,t)L(t)^{-3}$ vs. $kL(t)$, (e) $S_l(k,t)L_l(t)^{-3}$ vs. $kL_l(t)$ and (f) $S_H(k,t)L_H(t)^{-3}$ vs. $kL_H(t)$. The dashed line of slope -5 in (d) indicates the Porod tail characteristic of scattering off string defects. Similarly, the dashed lines with slope -4 in (e) and (f) indicate interfacial scattering.}}
\label{figure_7}
\end{figure}

We next focus on the scaling functions that describe the time-dependent morphologies. To evaluate them, we follow the same coarse-graining approach, but now the system is divided into non-overlapping sub-boxes of size ($6.0\sigma_0$)$^3$, which maps the system onto a simple cubic lattice of size ($16.0\sigma_0$)$^3$. This procedure gives us a continuous order parameter field and eliminates any molecular-level anisotropies. We have changed the system size to ensure that even at later times, the sub-boxes contain about 60 to 70 particles. In Fig.~\ref{figure_7}(a), we plot $C(r,t)$ vs. $r/L(t)$ for three specified values of $t$ and $\kappa^\prime$. The typical scalar nematic order parameter $\overline{P_2(\cos\theta_i,t)}$ is determined at each site $i$ of this discretized lattice (as for the nematic regime earlier) and subsequently the standard probes are evaluated. The data sets for $\kappa^\prime=5.0$ at different times neatly collapse onto a single master function, showing that the scaling regime has been reached. This data collapse indicates the existence of dynamical scaling. Furthermore, the excellent data collapse for different $\kappa^\prime$ values at time $t=98304$ suggests that the scaling functions are robust with respect to the anisotropy in energy. Fig.~\ref{figure_7}(d) shows the corresponding scaled structure factor, $S(k,t)L(t)^{-3}$ vs. $kL(t)$ for a range of $t$ and $\kappa^\prime$ values. There is an excellent data collapse, confirming both dynamical scaling and super-universality. The tail decays as $S(k,t)\sim k^{-5}$ due to the presence of string defects.

It is also possible to determine the correlation between the Sm layers by evaluating the translational correlation function $C_l(r,t)$. This can be done by evaluating the translational order parameter, ${\lvert\overline{\tau(\boldsymbol{r_i},t)}\rvert}={\lvert\overline{\exp(2i\pi\boldsymbol{r_i}\cdot\boldsymbol{n}/d_l)}\rvert}$. Here, $\boldsymbol{r_i}$ is the position vector of the LC molecule (located in the sub-box with index $i$), $\boldsymbol{n}$ is the global director of the system and the layer spacing $d_l$ is evaluated from the average separation between the peaks in the longitudinal pair correlation function at a given $t$. The overline implies an average over all the particles located in the sub-box with index $i$. Fig.~\ref{figure_7}(b) shows the scaled translational correlation function $C_l(r,t)$ vs. $r/L_l(t)$ for the specified values of $t$ and $\kappa^\prime$. The corresponding structure factors $S_l(k,t)L_l(t)^{-3}$ vs. $kL_l(t)$ are shown in Fig.~\ref{figure_7}(e). The high quality of the data collapse again confirms dynamical scaling and super-universality of the scaling functions. The structure factor tail exhibits the Porod decay, $S_l(k,t)\sim k^{-4}$, characteristic of scattering off sharp interfaces arising between the Sm layers. 

\textcolor{black}{ We have also determined the correlation function $C_H(r,t)$ using the bond-orientational order parameter ${\lvert\overline{\psi_6(\boldsymbol{r_i},t)}\rvert}$. Here, $\boldsymbol{r_i}$ is the position vector of the LC molecule located in the sub-box with index $i$, and the overline implies an average over all the particles located in the sub-box with index $i$. Fig.~\ref{figure_7}(c) shows the scaled correlation function $C_H(r,t)$ vs. $r/L_H(t)$ for the specified values of $t$ and $\kappa^\prime$. The corresponding structure factors $S_H(k,t)L_H(t)^{-3}$ vs. $kL_H(t)$ are shown in Fig.~\ref{figure_7}(f). The good data collapse re-confirms dynamical scaling and super-universality of the scaling functions.} 

\begin{figure}[ht]
\includegraphics[width=1.0\textwidth]{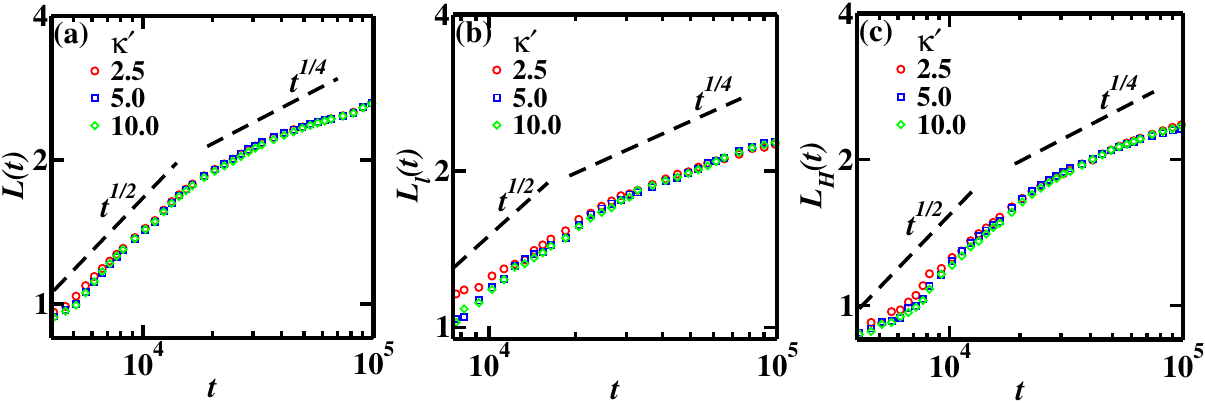}
\caption{\textcolor{black}{Growth laws (log-log scale) for the SmB-H phase at different values of $\kappa^\prime$: (a) $L(t)$ vs. $t$, (b) $L_l(t)$ vs. $t$  and (c) $L_H(t)$ vs. $t$.  Dashed lines of slope 1/2 and 1/4 are a guide to the eye. Early time data obeys the LAC law. The data indicates crossover to a slower growth at late times and hints at $\sim t^{1/4}$ growth. The two-time-scale scenario seen is reconfirmed by these distinct evaluations.}}
\label{figure_8}
\end{figure}

Finally, we study the domain growth laws for the SmB-H mesophase. Fig.~\ref{figure_8}(a) shows $L(t)$ vs. $t$\textcolor{black}{, Fig.~\ref{figure_8}(b) shows $L_l(t)$ vs. $t$ and Fig.~\ref{figure_8}(c) shows $L_H(t)$ vs. $t$} on a log-log scale for the three typical values of $\kappa^\prime$. (There are small differences in the pre-factors of the data sets, but they are not evident on the log-log scale.) The data sets exhibit an initial LAC growth regime $\sim t^{1/2}$ characteristic of the Nm phase and then a crossover to a slower $t^{1/4}$ growth regime. The dashed lines have been shown for reference. (Larger system sizes and longer simulation times will be required to remove the finite size effects observed at late times.)  These observations emphasize the two-time-scale scenario identified in Fig.~\ref{figure_6}, and are the second novel aspect of our study. 

\section{Summary and Discussion}
\label{summary} 

\begin{figure}[ht]
\includegraphics[width=1.0\textwidth]{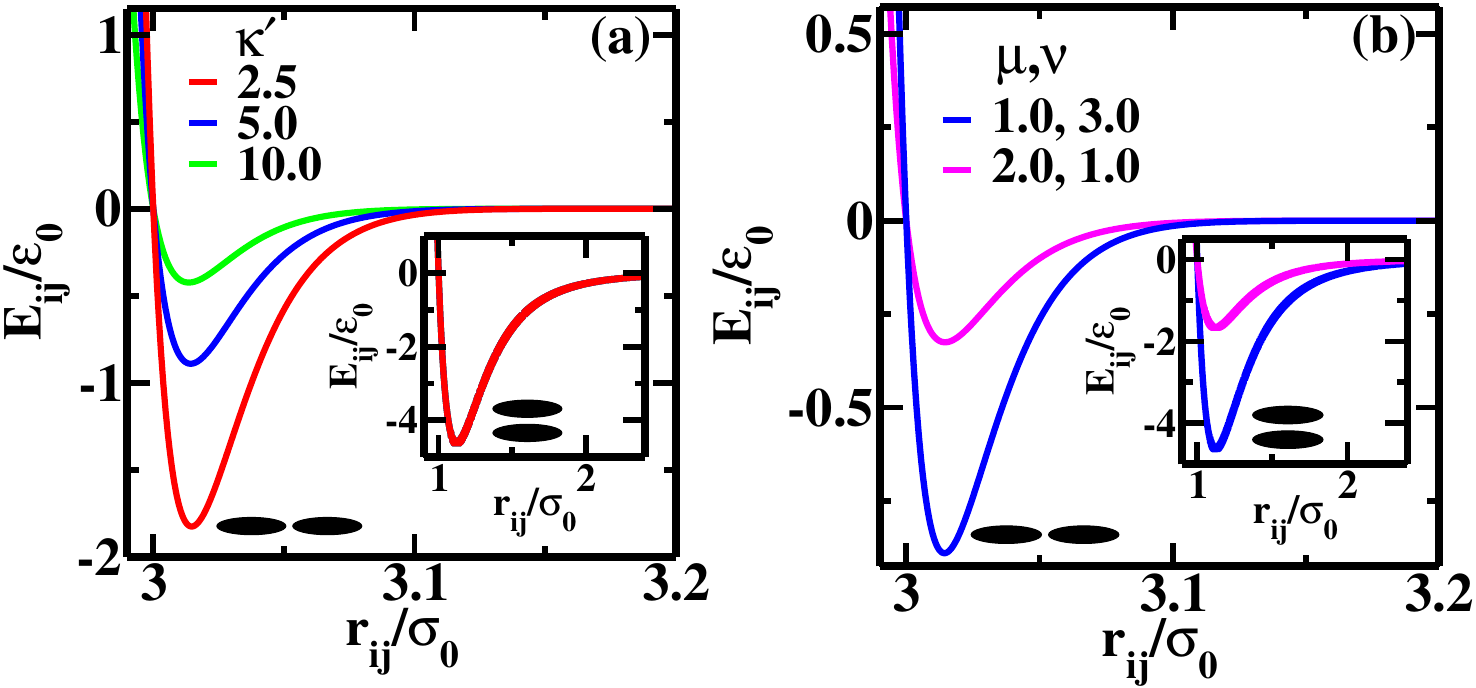}
\caption{(a) Variation of the scaled GB energy between a pair of ellipsoidal molecules as a function of their scaled separation ($E_{ij}/\epsilon_0$ vs. $r_{ij}/\sigma_0$) in the ee configurations with $\kappa=3.0$, $\mu=1.0$, $\nu=3.0$ for energy anisotropy $\kappa^\prime$ = 2.5 (red line), 5.0 (blue line) and 10.0 (green line). The inset shows the corresponding variation in the ss configuration. Notice that the curves for different values of $\kappa^\prime$ overlap. It can be observed that the energy anisotropy parameter affects the well depth for the ee configuration but not for the the ss configuration. (b) Similar variation for $\mu,\nu$ = 1.0, 3.0 (blue lines) and 2.0, 1.0 (magenta lines) with $\kappa=3.0$, $\kappa^\prime=5.0$. The exponents $\mu$ and $\nu$  affect the well depth in the ee as well as the ss configurations.}
\label{figure_1}
\end{figure}

We conclude with a summary and discussion of our results. We have undertaken a comprehensive numerical investigation of the equilibrium and non-equilibrium phenomena in the $d=3$ Gay-Berne model. This model is known to exhibit isotropic (I), nematic (Nm), and smectic (Sm) phases and yields satisfactory comparisons with experimental observations in liquid crystal systems. There are two important parameters in this model: (i) the shape anisotropy parameter $\kappa$ which is the length-to-breadth ratio of the ellipsoidal molecules; and (ii) the energy anisotropy parameter $\kappa^\prime$ which is the ratio of energies when molecules are in the side-by-side (ss) and in the end-to-end (ee) configurations, see Fig.~\ref{figure_1}. In all our studies, we make a standard choice of $\kappa=3.0$ and vary $\kappa^\prime$ over a wide range of values from 1.25 to 20. The primary focus of our work is to understand domain growth in the Sm mesophase. To the best of our knowledge, this is the first such study.

Equilibrium studies have been performed using canonical ($NVT$) ensemble {\it Monte Carlo} (MC) simulations on systems with 512 and 1000 particles. Equilibration was achieved in approximately $5\times10^5$ MC cycles and the observations were made in the window of $5\times10^5-10^6$ MC cycles. We confirmed the presence of I, Nm and Sm phases, and two distinct phase transitions: (i) I$\rightarrow$Nm at $T_c^{\ast 1}$ and (ii) Nm$\rightarrow$Sm at $T_c^{\ast 2}<T_c^{\ast 1}$. Our numerics indicate that $T_c^{\ast 1}$ decreases substantially as $\kappa^\prime$ increases, but $T_c^{\ast 2}$ increases only slightly.

\begin{figure}[ht]
\includegraphics[width=0.91\textwidth]{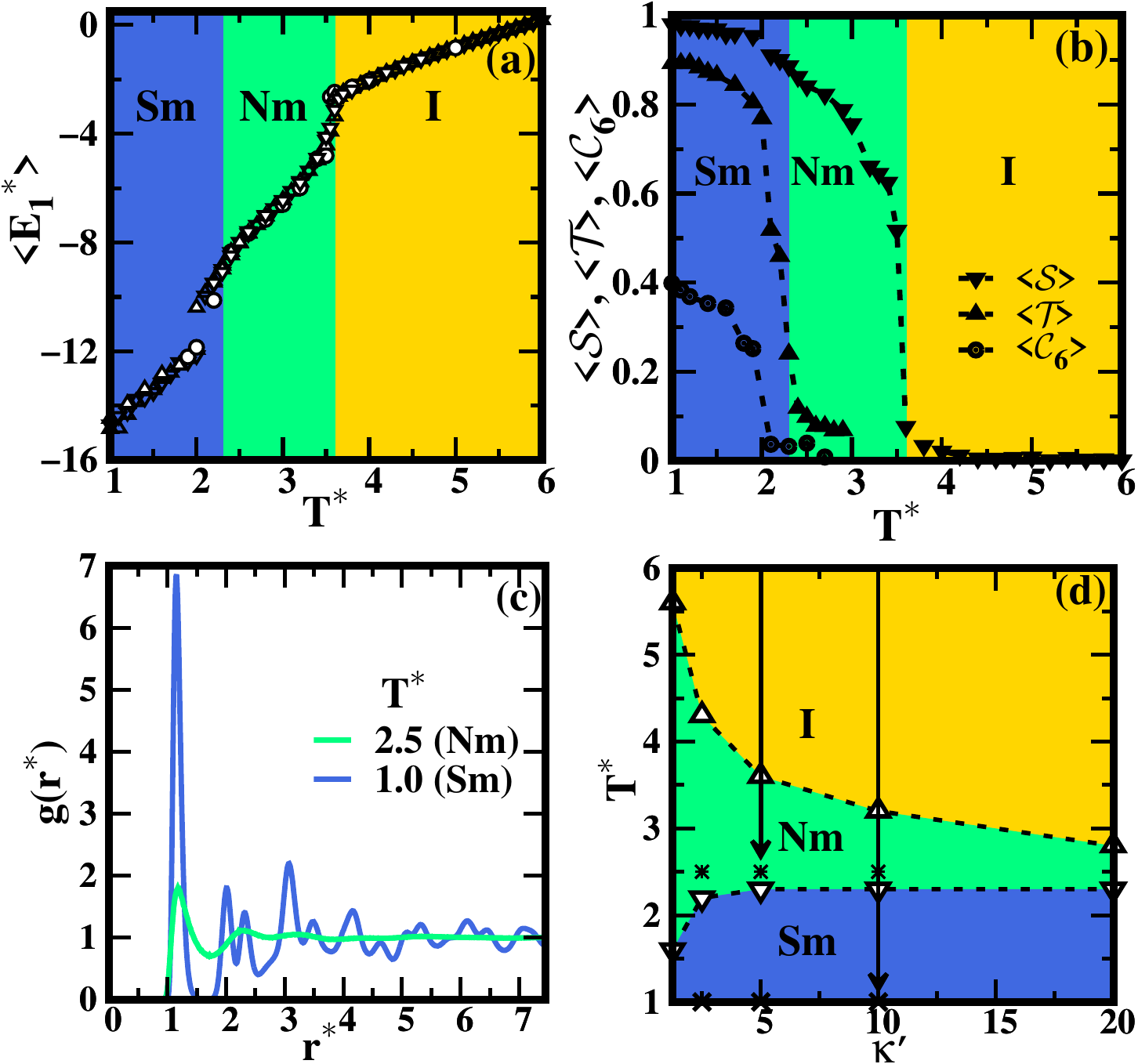}
\caption{(a) Variation of the scaled per-particle energy for the GB model system, $\langle E_1^\ast \rangle$ with respect to the scaled temperature $T^\ast$ with $\kappa=3.0$, $\kappa^\prime=5.0$, $\mu=1.0$ and $\nu=3.0$ for the two system sizes $N=512$ (open up triangles) and $N=1000$ (open down triangles). Corresponding results from Berardi {\it et al.} \cite{Berardi_1993} for $N=512$ (white circles), and for $N=1000$ during heating (white up triangles) and cooling (white down triangles) protocols are also shown for comparison. \textcolor{black}{(b) Variation of $\langle\mathcal{S}\rangle$ (scalar nematic), $\langle\mathcal{T}\rangle$ (translational) and $\langle\mathcal{C}_6\rangle$ (hexatic) order parameters with respect to $T^\ast$ at $\kappa^\prime=5.0$ for the $N=1000$ system. (c) Radial distribution function $g(r^\ast)$ vs. $r^\ast$ at $T^\ast$ = 1.0 (Sm phase) and $T^\ast$ = 2.5 ( Nm phase) for $\kappa^\prime=5.0$ and $N=1000$. (d) Phase diagram in the $T^\ast-\kappa^\prime$ plane indicating the isotropic (I) phase (yellow), nematic (Nm) phase (green) and Sm phase (blue). The asterisks and crosses indicate the quenches in the Nm and Sm regimes to study coarsening. The arrows indicate two typical quenches from I$\rightarrow$Nm and I$\rightarrow$Sm.} }
\label{figure_2}
\end{figure}

\begin{figure}[ht]
\includegraphics[width=0.76\textwidth]{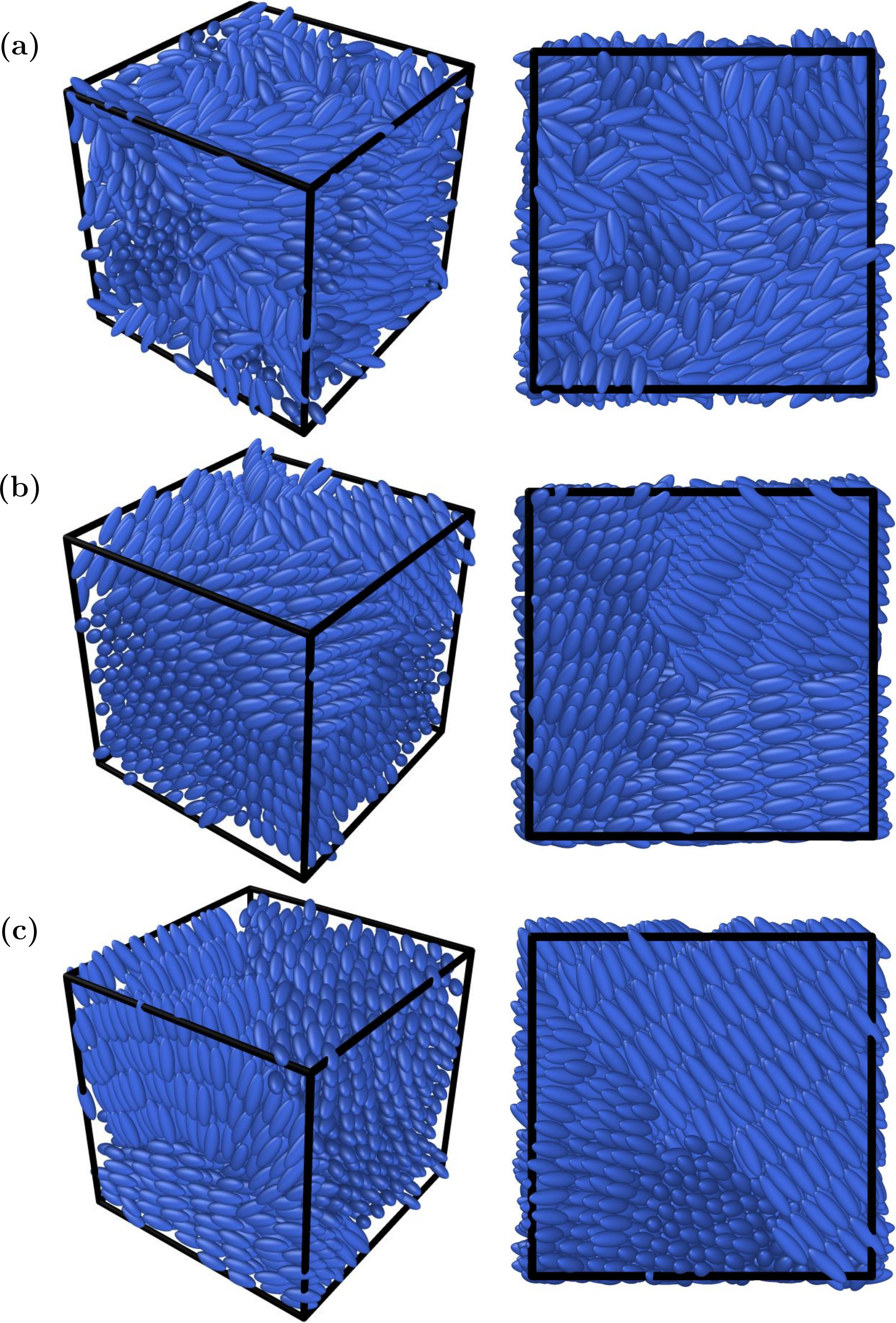}
\caption{Evolution snapshots after quenches from the I to the Sm phase [$T^\ast=1.0$, see Fig.\ref{figure_2}(d)] for (a) $\kappa^\prime=5.0$, $t=4096$; (b) $\kappa^\prime=5.0$, $t=98304$ and (c) $\kappa^\prime=10.0$, $t=98304$. Left: $20^3$ corner portions of the entire simulation box allowing for visibility of individual particles. Right: the corresponding top $d=2$ cross-sections. Notice that smecticity is enhanced by larger values of $\kappa^\prime$. \textcolor{black}{The layers are interdigitated and within a layer, the particles exhibit hexagonal order.}}
\label{figure_5}
\end{figure}

We studied the non-equilibrium phenomenon of coarsening via MD simulations of the Gay-Berne model with $\sim$260000 particles in the $NVT$ ensemble. An initially disordered and homogeneous state was rapidly quenched from (a) I$\rightarrow$Nm and (b) I$\rightarrow$Sm phases, see Fig.~\ref{figure_2}(d). The system was then allowed to evolve till late times, and we identified the morphology textures and growth laws through this evolution. \\
(a) Our results for the nematic quench (I$\rightarrow$Nm) are as follows: the equal-time spatial correlation function $C(r,t)$ exhibits dynamical scaling and is robust with respect to $\kappa^\prime$. Coarsening is due to annihilation of the string defects [$S(k,t)\sim k^{-5}$], and the domain growth obeys the {\it Lifshitz-Allen-Cahn} (LAC) law: $L(t)\sim t^{1/2}$. Earlier results in the literature were inconclusive due to small system sizes used in simulations \cite{Billeter_1999}. \\
(b) In our I$\rightarrow$Sm quenches, the low temperature phase has a Sm B hexatic (SmB-H) order.  With regard to our novel results on coarsening in this phase, the domain growth exhibits a two-time-scale scenario: With the onset of coarsening, the LC molecules align and develop orientational order (nematicity). \textcolor{black}{The arrangement in layers (smecticity) with hexatic bond-orientational-order (BOO) within the layers follows thereafter. Consequently, the growth follows the LAC law, $L(t)\sim t^{1/2}$ at early times and then a crossover to a slower growth at later times. Interestingly, the correlation functions and corresponding structure factors show dynamical scaling in both the regimes with universal scaling functions. These are also robust with respect to $\kappa^\prime$, and the smecticity and BOO are pronounced at larger values.} Furthermore, the early-time dynamics is governed by string defects [$S(k,t)\sim k^{-5}$], while the late-time evolution is dictated by interfacial defects [$S_l(k,t)\sim k^{-4}$ \textcolor{black}{and $S_H(k,t)\sim k^{-4}$}]. We believe these results to be valid for other classes of the Sm phase as well. 

In conclusion, we believe that the novel results presented in this paper reveal many unusual aspects of ordering in the Sm phase. Although of consequence in the LC arena, it has received little attention due to experimental and computational difficulties. The methodology presented in this work to study equilibrium and non-equilibrium phenomena in calamitic LCs could also be applied to study the mesophases that occur in discotic and bent-core LC systems. Another significant aspect of our work has been incorporation of the Gay-Berne potential in the general-purpose MC program,  DL\_MONTE \cite{DL_Monte}. This new functionality should prove useful for future MC studies  of uniaxial as well as biaxial and bent core LC systems.  

\acknowledgments{NB acknowledges UGC, India for a senior research fellowship. VB acknowledges DST, India for a core research grant. NB, TU, NBW and VB also acknowledge DST-UKIERI for a research grant which has facilitated this collaboration. NB and TU acknowledge Prof. Steve Parker at the University of Bath (UK) for the kind hospitality provided during the developmental stages of this work. NB and VB gratefully acknowledge the High Performance Computing (HPC) facility at IIT Delhi for computational resources.}

\bibliography{Ref}

\newpage 

\end{document}


\title{Supplementary Material\\
Equilibrium phases and domain growth kinetics of calamitic liquid crystals}

\author{Nishant Birdi}
\email{srz188382@sire.iitd.ac.in}
\affiliation{School of Interdisciplinary Research, Indian Institute of Technology, Hauz Khas, New Delhi 110016, India.}

\author{Tom L. Underwood}
\email{tlu20@bath.ac.uk}
\affiliation{Department of Chemistry, University of Bath, Bath BA2 7AY, United Kingdom.}

\author{Nigel B. Wilding}
\email{nigel.wilding@bristol.ac.uk}
\affiliation{H.H. Wills Physics Laboratory, University of Bristol, Royal Fort, Bristol BS8 1TL, United Kingdom.}

\author{Sanjay Puri}
\email{purijnu@gmail.com}
\affiliation{School of Interdisciplinary Research, Indian Institute of Technology, Hauz Khas, New Delhi 110016, India.}
\affiliation{School of Physical Sciences, Jawaharlal Nehru University, New Delhi 110067, India.}

\author{Varsha Banerjee}
\email{varsha@physics.iitd.ac.in}
\affiliation{School of Interdisciplinary Research, Indian Institute of Technology, Hauz Khas, New Delhi 110016, India.}
\affiliation{Department of Physics, Indian Institute of Technology, Hauz Khas, New Delhi 110016, India.} 


\maketitle 

\section{Improvements to DL\_MONTE}
In DL\_MONTE the centers of interaction in the system are termed \emph{atoms}. Atoms in DL\_MONTE can interact with one another in a number of ways, though the
most significant is through pair potentials. Note, however, that before this work only \emph{isotropic} pair potentials were supported in DL\_MONTE.

DL\_MONTE allows for the possibility to construct \emph{molecules} comprised of two or more atoms, whose relative positions within
the molecule are constrained, and to treat the positions and orientations of the molecules as the degrees of freedom of the system instead
of the atoms' positions. By doing this, a number of well studied liquid-crystal system models, models in which the particles have orientational
degrees of freedom and \emph{anisotropic} interactions between particles, can be accommodated in DL\_MONTE. For instance, in the rigid linear fused hard-sphere chain
model \cite{Jaffer_1999} each particle is comprised of $n$ overlapping hard spheres of radius $\sigma$ residing on a line in space, spaced at length $d<\sigma$ apart from
one another. This model could be realized in DL\_MONTE by constructing a system of molecules comprised of $n$ atoms per molecule, with the 
positions of the atoms in each molecule conforming to the constraints just described, and specifying that all atoms interact with atoms in other molecules via a hard-sphere potential with radius $\sigma$.
However, there exist many models pertinent to liquid crystals which cannot be treated in this way. To elaborate, the aforementioned approach of building molecules from atoms to create anisotropic particles only
works for models in which the interactions between particles can be expressed as a \emph{sum of isotropic pair potentials} emanating from one or more
`atoms' within the particles. For instance, the GB model cannot be treated this way, since in the GB model each particle has only one centre of
interaction, i.e. one atom, and the strength of the interaction between two particles $i$ and $j$ depends not only on the separation vector between the
atoms $\mathbf{r}_{ij}$, but also on implicit orientations $\mathbf{u}_i$ and $\mathbf{u}_j$ associated with the atoms.

To enable DL\_MONTE to accommodate GB model, as well as other models which describe systems of \emph{linear} particles (e.g. hard ellipsoid, hard
spherocylinder), for which the form of the potential between two particles is $V_{ij}(\mathbf{r}_{ij},\mathbf{u}_i,\mathbf{u}_j)$, we made the following
fundamental modifications to the code.
Firstly, we extended the atom data type to include an orientational degree of freedom analogous to $\mathbf{u}_i$ or $\mathbf{u}_j$ above. Secondly, we generalized the underlying form of the pair potential between two atoms from $V_{ij}(\mathbf{r}_{ij})$ to
$V_{ij}(\mathbf{r}_{ij},\mathbf{u}_i,\mathbf{u}_j)$. Thus atoms could represent \emph{linear} particles interacting with orientation-dependent pair potentials.
Finally, we introduced a new MC move into DL\_MONTE for modifying an atom's orientation -- something necessary for
sampling the orientational degrees of freedom in a system. This MC move uses quaternions to rotate the orientation $\mathbf{u}_i$ of an atom $i$, chosen at random,
about a
randomly chosen axis by an angle drawn uniformly from within the range $[-\theta_{\text{max}},\theta_{\text{max}}]$, where $\theta_{\text{max}}$ is the
maximum possible angle of rotation \cite{Frenkel_2002}. Moreover, for the convenience of users, we also added functionality to enable DL\_MONTE to optimize
$\theta_{\text{max}}$ during a simulation to yield a desired acceptance rate. (Similar functionality exists in DL\_MONTE for other MC move types).
These fundamental changes to DL\_MONTE were accompanied by further additions to the code geared towards studying the GB system specifically. One obvious
change was to implement the GB potential. Another was to enable DL\_MONTE to calculate and output the scalar nematic order
parameters $\langle P_2(\mathbf{u}_i\cdot\mathbf{n})\rangle$ and $\langle P_4(\mathbf{u}_i\cdot\mathbf{n})\rangle$ periodically during the simulation, 
where $P_n$ denotes the $n$th Legendre polynomial, $\mathbf{n}$ is the director of the system, and $\langle\rangle$ denotes an
ensemble average.
We exploited this feature during this work to characterize the phase of the system at various
thermodynamic parameters -- as described in the main text.
These improvements make DL\_MONTE a versatile tool for studying off-lattice liquid crystal models with MC simulation.

\section{Input Files for MC simulations using DL\_MONTE}
\label{dlmonte}

The program DL$\textunderscore$MONTE requires three necessary input files for any simulation, namely CONTROL, CONFIG and FIELD. These three files should always be present in the directory where the simulation is run.\\
(i) The CONTROL file provides the general input parameters for a simulation run, which are required to instruct DL$\textunderscore$MONTE on how to undertake the calculations. These parameters contain information about the thermodynamic conditions and statistical ensembles in which the simulations are performed, thermal averaging, sampling schemes, output frequencies, format of output files in which the simulation data is to be stored, etc.\\
(ii) The CONFIG file provides the starting configuration for the simulation. It contains detailed specifications for the simulation cell (box) such as its dimensions and geometry, and the initial microscopic state in terms of the position coordinates for all the particles. For particles interacting with the recently incorporated GB potential, the initial state is specified with their respective orientations in addition to positions.\\
(iii) The FIELD file provides the atomic and/or the molecular species specifications. These include information about their general topology i.e. intramolecular bonds, angles, etc. and the force fields (interaction potentials).\\ 
Below we provide these three files with all the commands and parameters (with necessary comments after the \# symbol), for MC simulation of the $\kappa=3.0$, $\kappa^\prime=5.0$, $\mu=1.0$ and $\nu=3.0$ GB system with $N=1000$ at $T=1.0$ respectively. Note that we have numbered all the commands only for a clear and better understanding of the reader, the numbering is not required in the actual input files.
Further information regarding DL\_MONTE, including the input files and how to install the program, can be found elsewhere \cite{Brukhno_2019,DL_Monte}.\\

\textbf{CONTROL file}\\
(1) NVT simulation of GB potential \# simulation title (must be within 80 characters)\\
(2) use ortho \# `use' section begins; `ortho' means to use orthogonal PBC, applicable for cubic and orthorhombic cells\\
(3) finish \# keyword to close the `use' section\\
(4) seeds 30 4 19 94 \# enter any 4 `seed' numbers required for random number generation\\               
(5) nbrlist auto \# update the neighbourlist (nl) using `auto' mode \\                 
(6) maxnonbondnbrs 1000 \# maximum number of non-bonded neighbors in nl set to 1000 \\              
(7) temperature    1.0 \# set the system temperature to 1.0\\
(8) steps          1000000000 \# specify total number of MC moves in the simulation\\          
(9) equilibration  0 \# set the equilibration period to 0 (we select the period once the simulation is over)\\    
(10) print          1000000 \# print data to the OUTPUT file after the specified no. of moves  \\                    
(11) revconformat  vtk \# `vtk' format for the REVCON output file         \\ 
(12) archiveformat vtk \# `vtk' format for the trajectory file         \\                       
(13) sample coords 100000000 \# archive snapshots in trajectory file after the specified no. of moves\\
(14) sample rdf 745 7.45 1000000 \# compute $g(r^\ast)$ with 500 bins, upto $r^\ast=5.0$, collecting data after every $10^6$ moves\\
(15) maxatmrot 15 \# maximum angle (in degrees) to rotate a LC molecule during the MC move\\
(16) acceptatmrotupdate 1000000010 \# frequency of updating the maximum angle to rotate a LC molecule: we set a value so that the update does not happen as it's not required in our simulation\\
(17) acceptatmrotratio 0.5 \# target acceptance ratio set to 0.5\\
(18) move atom 1 100 \# attempt translating atoms of type `1' (all GB particles) at frequency of `100' (i.e. 100\% of the time) during the simulation \\
(19) GB spin \# translate atom of name `GB' and type `spin' during the simulation\\
(20) yamldata scalarop2 \# output $\langle P_2 \rangle$ order parameter to the YAMLDATA output file\\
(21) yamldata scalarop4 \# output $\langle P_4 \rangle$ order parameter to the YAMLDATA output file \\
(22) yamldata 10000 \# print data to the YAMLDATA output file after the specified no. of moves\\
(23) stat 1000000000 \# print statistics to the PTFILE after the specified no. of moves\\
(24) move rotateatom 1 100 \# attempt rotating atoms of type `1' at frequency `100' during the simulation \\
(25) GB spin \# rotate atom of name `GB' and type `spin' during the simulation\\
(26) start   \# this command indicates that all the required inputs have been given and now `start' the simulation\\

\textbf{CONFIG file:}\\
(1) GB Potential for Nematic LC in NVT ensemble \# configuration title \\
(2) 0 0 \# The first integer here is set to 0 as for all the particles, only the position coordinates are specified in the file (non-zero integer values indicate that in addition to their position coordinates, velocities and/or forces on particles are to be specified). The second integer indicates that the particle positions are stored in {\it fractional} coordinates and not Cartesian coordinates (`1' for cubic, `2' for orthorhombic and `3' for parallelepiped). These are then followed by specification of the cell matrix in the next lines.\\
\# The next 3 lines indicate the $x$, $y$ and $z$ components for the 3 simulation cell vectors:
(3) 14.938 0.0 0.0\\
(4) 0.0 14.938 0.0\\
(5) 0.0 0.0 14.938\\
(6) NUMMOL 1 1 \# specifies the no. of types of molecules followed by no. of each type: in our case, only 1 type of molecule is present in the simulation\\
(7) MOLECULE gb 1000 1000 \# 1 molecule of type `gb': the molecule has 1000 particles initially and is limited to a maximum number of 1000\\
\# Now for all the 1000 particles (atoms), their name `GB' and type `spin' (first line), 3 fractional position coordinates (second line) and 3 orientation components (third line) are to be specified successively for all the particles.\\   
(8) GB spin\# particles (atoms) have name `GB' and type `spin'\\
(9) 0 0 0 \# fractional position coordinates for the first atom \\ 
(10) 0.577350269 0.577350269 0.577350269 \# orientation components for the first atom \\
(11) GB spin\\
(12) 0 0 0.1 \# fractional position coordinates for the second atom \\
(13) 0.577350269 0.577350269 0.577350269 \# orientation components for the second atom \\
(14) GB spin\\
(15) 0 0 0.2 \# fractional position coordinates for the third atom \\
(16) 0.577350269 0.577350269 0.577350269 \# orientation components for the third atom \\
(17) GB spin\\
(18) 0 0 0.3 \# fractional position coordinates for the fourth atom \\
(19 )0.577350269 0.577350269 0.577350269 \# orientation components for the fourth atom \\
..... \# and so on for all the atoms\\ 

\textbf{FIELD file:}\\
(1) Gay-Berne, 4\*sigma\_s cut-off, sigma\_s = 1 Angstrom, epsilon\_0 = 1 K \# force-field title \\
(2) CUTOFF 4 \# shortrange cut-off radius for interactions\\
(3) UNITS K \# energy unit for input and output: `K'$\equiv k_BT$ units\\
(4) NCONFIGS 1 \# number of configuration boxes (replicas) in the CONFIG file set to 1\\
(5) ATOM TYPES 1 \# only 1 type of atoms are present in our simulation (which are the GB types)\\
(6) GB spin 1.0 0.0 \# atom name is `GB' (user defined name) with new type `spin' introduced in DL\_MONTE (meaning particles with orientations), mass set to 1.0 and charge set to 0 (these atoms are contained within a single structureless molecule called `gb')\\
(7) MOLECULE TYPES 1 \# only 1 type of molecules present in our simulation\\
(8) gb \# molecule name is `gb' (user defined name)\\
(9) MAXATOM 1000 \# maximum number of atoms for this type of molecule (the number is set to the total number of particles in our simulation $N$)\\
(10) FINISH \#  close the molecule specification section\\
(11) VDW 1 \# non-bonded VdW section starts with the number of participating atom type pairs, which is only 1 in our case \\
(12) GB spin GB spin gb 1.0 5.0 1.0 3 1 3 \# The 2 interacting atoms have name `GB', type `spin' and they interact through the `gb' (Gay-Berne) interaction with parameters specified in the given order: $\epsilon_0=1.0$, $\kappa^\prime=5.0$, $\sigma_0=1.0$, $\kappa=3.0$, $\mu=1.0$ and $\nu=3.0$.\\ 
(13) CLOSE \# close the FIELD file after all the necessary inputs have been given\\

\section{Input Files for MD simulations using LAMMPS}
\label{lammps}

Unlike DL\_MONTE, for LAMMPS we require only a single input file with all the necessary commands. Our input files (with necessary comments) to prepare the initial homogeneous disordered state and then quench the system to the nematic state, are presented below separately. Again we have numbered all the commands only for a clear and better understanding of the reader, the numbering is not required in the actual input files.\\  

\textbf{Lammps input file to prepare the homogeneous disordered system:}\\
\# Initialization \\
(1) units lj \# style of units used in this simulation is `lj' in which all the quantities are unitless\\
(2) newton on \# turn Newton's third law `on' for pairwise and bonded interactions\\ 
(3) dimension 3 \# dimensionality of the simulation is set to 3\\
(4) processors 4 4 4 \# 4 processors in each dimension of the 3d grid overlaying the simulation domain, hence total 4x4x4=64 processors used in this simulation (command needed for parallel and efficient computing)\\  
(5) boundary p p p \# set the style of boundaries for the global simulation box in each dimension as `periodic' (p)\\
\# Setup the Simulation Box and the Atoms \\
(6) atom\_style ellipsoid \# set the style of atoms used in this simulation as `ellipsoid'\\
(7) lattice sc 0.3 \# define a lattice which is of style `simple cubic' (sc), reduced density of the system is set to 0.3\\
(8) region my\_box block 0.0 64.0 0.0 64.0 0.0 64.0 \# total number of particles in the system is 64x64x64, 'my\_box' is the user-assigned name for the `block' style region\\ 
(9) create\_box 1 my\_box \# create a simulation box based on the region specified with the user-defined name, only 1 type of particles are present in this simulation\\  
(10) create\_atoms 1 box \# creates atoms on the lattice defined earlier\\  
(11) set type 1 mass 1.0 \# set the mass of all the atoms to unity\\
(12) set type 1 shape 1.0 1.0 3.0 \# set the 3 diameters of the ellipsoids to 1.0, 1.0 and 3.0\\ 
(13) set type 1 quat 0.5773502692 0.5773502692 0.5773502692 45 \# set the quaternions which represent orientation of ellipsoids: the first 3 numbers are for the unit vector to rotate the ellipsoid around via the right-hand rule and the last number is for the rotation angle (in degrees)\\
(14)velocity all create 6.0 250964 mom yes rot yes loop geom \# set velocities for the atoms by generating an ensemble of velocities using a random number generator with the specified seed number (i.e. 250964) at the specified temperature (i.e. 6.0), `mom yes' sets the linear momentum of the ensemble to zero, `rot yes' sets the angular momentum of the ensemble to zero, and `loop geom' is a command for parallelization using which each processor loops over only its atoms.\\
\# Force fields (Potential) \\
(15) pair\_style gayberne 1.0 3.0 1.0 4.0 \# set the formula used for pairwise interactions which is the GB model in our case, followed by the required parameters: shift for potential minimum (set to 1.0), exponent $\nu$ (set to 3.0), exponent $\mu$ (set to 1.0) and the global cut-off for interactions (set to 4.0)\\  
(16) pair\_coeff 1 1 1.0 1.0 1.0 1.0 0.2 1.0 1.0 0.2 4.0 \# set the model parameters in the following order: the first two numbers indicate the types of the interacting atoms between which the potential is to be computed, the third number is the well depth (in energy units), the fourth number is the minimum effective particle radii (in distance units), the next three numbers indicate the relative well depth for the side-to-side, face-to-face and end-to-end interactions of the first particle followed by the relative well depth for same interactions of the second particle and the last number is the cut-off\\ 
(17) comm\_style brick \# the `brick' style domain decomposition is used during parallelization\\
(18) comm\_modify mode single vel yes\# during domain decomposition, atoms are communicated within a single cut-off distances and velocity information is also communicated\\
(19) neighbor 1.0 bin \# set parameters to build the pairwise neighbor lists; all the atom pairs within a neighbor cut-off distance equal to their force cut-off plus the skin distance (set to 1.0 here) are stored in the list created using the `bin' style where binning scales linearly with the number of atoms per processor\\
(20) neigh\_modify every 1 delay 0 check yes \# set parameters that affect the building and use of pairwise neighbor lists: `every 1' means build the list after every step, `delay 0' means delay building until 0 steps since the last build and `check yes' means to build only if some atom has moved half the skin distance or more\\
\# Output Settings \\
(21) thermo 100 \# output the thermodynamics after every 100 time-steps\\
(22) timestep 0.001 \# set the MD timestep size to 0.001\\
(23) compute q all property/atom quatw quati quatj quatk \# compute the 4 components of quaternions representing orientations of the atoms\\    
(24) compute sh all property/atom shapex shapey shapez \# compute the 3 shape parameters (diameters of the ellipsoids) for all the atoms\\
(25) dump 1 all custom 50000 dump.GayBerne id type x y z vx vy vz c\_q[1] c\_q[2] c\_q[3] c\_q[4] c\_sh[1] c\_sh[2] c\_sh[3] \# store the system configuration in a dump file after every `50000' time-steps, where the position coordinates, velocity coordinates, quaternion components for orientations and the 3 shape parameters are specified for all the atoms at a given timestep\\  
(26) group uniaxial type 1 \# make a group with name `uniaxial' consisting of all the type 1 particles\\
(27) variable dof equal count(uniaxial) \# `dof' is a variable with value equal to the number of particles in the `uniaxial' group\\
(28) fix 1 all nvt/asphere temp 6.0 6.0 0.1 \# perform constant $NVT$ integration to update the position, velocity, orientation and angular velocity at every timestep for the ellipsoidal particles using a Nos\'e-Hoover temperature thermostat and hence create a system trajectory consistent with the canonical ensemble, the three numbers are external temperature at start of the run, external temperature at end of the run and temperature damping parameter (in time units) which determines how rapidly the temperature is relaxed\\
(29) compute\_modify 1\_temp extra/dof \${dof} \# subtract dof degrees-of-freedom as a normalizing factor in a temperature computation\\
(30) thermo\_style custom step temp press c\_1\_temp evdwl epair pe ke etotal vol density lx ly lz nbuild ndanger \# set the style and content for printing thermodynamic data to the screen and log file, the output thermodynamic data in our case includes the following: temperature, pressure, potential energy, kinetic energy, total energy, volume, density, simulation box dimensions and number of neighbor list builds\\ 
(31) restart 50000 GB.dat \# write a restart file after every `50000' time-steps, `GB.dat' is the filename to which data at a given timestep is appended\\
(32) run\_style verlet \# the standard velocity-Verlet integrator is used\\
(33) run 1000000 \# run dynamics for the specified number of time-steps\\

\textbf{Input file to quench the system from high temperature disordered state to the nematic state:} (for same commands, we do not put comments again)\\
(1) processors 4 4 4\\
(2) read\_restart GB.dat.100000 \# read the given restart file which contains the initial high temperature configuration\\
\# Force fields (Potential)\\ 
(3) pair\_style gayberne 1.0 3.0 1.0 4.0\\
(4) pair\_coeff 1 1 1.0 1.0 1.0 1.0 0.2 1.0 1.0 0.2 4.0\\
(5) comm\_style brick\\
(6) comm\_modify mode single vel yes\\
(7) neighbor 1.0 bin\\
(8) neigh\_modify every 1 delay 0 check yes\\
\# Output Settings \\
(9) timestep 0.001\\
(10) reset\_timestep 0 \# set the timestep counter to the specified value\\
(11) compute q all property/atom quatw quati quatj quatk\\
(12) compute sh all property/atom shapex shapey shapez\\
\# NVT Run\\ 
(13) group uniaxial type 1\\
(14) variable dof equal count(uniaxial)\\
(15) fix 1 all nvt/asphere temp 2.5 2.5 0.1\\
(16) compute\_modify 1\_temp extra/dof \${dof}\\
(17) thermo\_style custom step temp press c\_1\_temp evdwl epair pe ke etotal vol density lx ly lz nbuild ndanger\\
(18) restart 100000 GB.dat\\
(19) run\_style verlet\\
(20) thermo 4\\
(21) dump 1 all custom 4 dump.GayBerne\_1 id type x y z vx vy vz c\_q[1] c\_q[2] c\_q[3] c\_q[4] c\_sh[1] c\_sh[2] c\_sh[3]\\
(22) run 32 upto\\
(23) undump 1\\
(24) thermo 8\\
(25) dump 2 all custom 8 dump.GayBerne\_2 id type x y z vx vy vz c\_q[1] c\_q[2] c\_q[3] c\_q[4] c\_sh[1] c\_sh[2] c\_sh[3]\\
(26) run 96 upto\\
(27) undump 2\\
(28) dump 3 all custom 16 dump.GayBerne\_3 id type x y z vx vy vz c\_q[1] c\_q[2] c\_q[3] c\_q[4] c\_sh[1] c\_sh[2] c\_sh[3]\\
(29) run 224 upto\\
(30) undump 3\\
(31) dump 4 all custom 32 dump.GayBerne\_4 id type x y z vx vy vz c\_q[1] c\_q[2] c\_q[3] c\_q[4] c\_sh[1] c\_sh[2] c\_sh[3]\\
(32) run 512 upto\\
(33) undump 4\\
(34) dump 5 all custom 64 dump.GayBerne\_5 id type x y z vx vy vz c\_q[1] c\_q[2] c\_q[3] c\_q[4] c\_sh[1] c\_sh[2] c\_sh[3]
(35) run 1024 upto
(36) undump 5\\
(37) dump 6 all custom 128 dump.GayBerne\_6 id type x y z vx vy vz c\_q[1] c\_q[2] c\_q[3] c\_q[4] c\_sh[1] c\_sh[2] c\_sh[3]\\
(38) run 2048 upto\\
(39) undump 6\\
(40) dump 7 all custom 256 dump.GayBerne\_7 id type x y z vx vy vz c\_q[1] c\_q[2] c\_q[3] c\_q[4] c\_sh[1] c\_sh[2] c\_sh[3]\\
(41) run 4096 upto\\
(42) undump 7\\
(43) dump 8 all custom 512 dump.GayBerne\_8 id type x y z vx vy vz c\_q[1] c\_q[2] c\_q[3] c\_q[4] c\_sh[1] c\_sh[2] c\_sh[3]\\
(44) run 8192 upto\\
(45) undump 8\\
(46) dump 9 all custom 1024 dump.GayBerne\_9 id type x y z vx vy vz c\_q[1] c\_q[2] c\_q[3] c\_q[4] c\_sh[1] c\_sh[2] c\_sh[3]\\
(47) run 16834 upto\\
(48) undump 9\\
(49) dump 10 all custom 2048 dump.GayBerne\_10 id type x y z vx vy vz c\_q[1] c\_q[2] c\_q[3] c\_q[4] c\_sh[1] c\_sh[2] c\_sh[3]\\
(50) run 32768 upto\\
(51) undump 10\\
(52) dump 11 all custom 4096 dump.GayBerne\_11 id type x y z vx vy vz c\_q[1] c\_q[2] c\_q[3] c\_q[4] c\_sh[1] c\_sh[2] c\_sh[3]\\
(53) run 65536 upto\\
(54) undump 11\\
(55) dump 12 all custom 8192 dump.GayBerne\_12 id type x y z vx vy vz c\_q[1] c\_q[2] c\_q[3] c\_q[4] c\_sh[1] c\_sh[2] c\_sh[3]\\
(56) run 98304 upto\\
(57) undump 12\\
(58) thermo 100\\
(59) run 100000 upto\\

\bibliography{Ref}